\documentclass[final]{ieeetran}

\usepackage[utf8]{inputenc}

\usepackage{stmaryrd} 
\usepackage{cite} 

\usepackage{amsmath}
\usepackage{amssymb}
\usepackage{amsfonts}
\usepackage{amscd}
\usepackage{mathtools}  

\usepackage{tabularx}

\usepackage{pgf,tikz}  
\usetikzlibrary{arrows} 
\usepackage{tikz-cd}  

\newcommand{\Ex}[1]{\mathbb{E}\left(#1\right)}
\renewcommand{\Pr}[1]{\mathbb{P}\left(#1\right)}
\newcommand{\Prr}[2]{\mathbb{P}_{#1}\left(#2\right)}
\newcommand{\Var}{\mathbin{\mathbb{V}}}

\newcommand{\qBin}{\operatorname{Bin}_q}
\newcommand{\Bin}{\operatorname{Bin}}
\newcommand{\Ber}{\operatorname{Ber}}

\newcommand{\diff}[2]{\frac{\mathrm{d}  #1}{\mathrm{d}  #2}}

\newcommand{\difp}[2]{\frac{\partial #1}{\partial #2}}

\def \d{\mbox{\(\,\mathrm{d}\)}}
 
\renewcommand{\epsilon}{\varepsilon} 
\DeclarePairedDelimiter\ceil{\lceil}{\rceil}
\DeclarePairedDelimiter\floor{\lfloor}{\rfloor}

\newcommand{\sm}{\smallsetminus}

\newcommand{\Rr}{\mathbb{R}}
\newcommand{\Cc}{\mathbb{C}}
\newcommand{\Nn}{\mathbb{N}}

\newcommand{\Zz}{\mathbb{Z}}
\newcommand{\Ee}{\mathbb{E}}
\newcommand{\Ff}{\mathbb{F}}

\newcommand{\set}[2]{\{\,#1 \, : \, #2\,\} }

\newcommand{\qbinom}[2]{ {#1 \brack #2}_q }
\renewcommand{\v}[1]{\mathbf {#1}}
\newcommand{\Grass}{\operatorname{Grass}}
\newcommand{\Dil}{\operatorname{Dil}}
\newcommand{\Gr}{\operatorname{Gr}}

\newcommand{\ent}[1]{H_{#1}}
\newcommand{\e}{\mathrm{e}}

\newtheorem{theorem}{Theorem}
\newtheorem{definition}{Definition}
\newtheorem{proposition}{Proposition}
\newtheorem{lemma}{Lemma}
\newtheorem{corollary}{Corollary}
\newtheorem{remark}{Remark}

\begin{document}
\title{Information theory with finite vector spaces}
\author{Juan Pablo Vigneaux
\thanks{Manuscript received August 28, 2018; revised February 8, 2019; accepted March 7, 2019. This paper was presented in part at the Latin American Week on Coding and Information 2018 (Campinas, Brazil).

Juan Pablo Vigneaux is with the {Institut de Math\'ematiques de Jussieu-Paris Rive Gauche}, attached administratively to Universit\'e Paris Diderot, F-75013 Paris, France; to Sorbonne Universit\'e, F-75005 Paris, France, and also to CNRS, F-75016 Paris, France (e-mail:juan-pablo.vigneaux@imj-prg.fr).

Communicated by P. Harremo\"es, Associate Editor for Probability and Statistics.

Digital Object Identifier 10.1109/TIT.2019.2907590

Copyright (c) 2019 IEEE. Personal use of this material is permitted. However, permission to use this material for any other purposes must be obtained from the IEEE by sending a request to pubs-permissions@ieee.org.
}}

\maketitle

\begin{abstract}
Whereas Shannon entropy is related to the growth rate of multinomial coefficients, we show that the quadratic entropy (Tsallis 2-entropy) is connected to their $q$-deformation; when $q$ is a prime power, these $q$-multinomial coefficients count flags of finite vector spaces with prescribed length and dimensions. In particular, the $q$-binomial coefficients count vector subspaces of given dimension. We obtain this way a combinatorial explanation for the nonadditivity of the quadratic entropy, which arises from a recursive counting of flags. We show that statistical systems whose configurations are described by flags provide a frequentist justification for the maximum entropy principle with Tsallis statistics. We introduce then a discrete-time stochastic process associated to the $q$-binomial probability distribution, that generates at time $n$ a vector subspace of $\mathbb{F}_q^n$ (here $\Ff_q$ is the finite field of order $q$). The concentration of measure on certain ``typical subspaces" allows us to extend the asymptotic equipartition property to this setting. The size of the typical set is quantified by the quadratic entropy. We discuss the applications to Shannon theory, particularly to source coding, when messages correspond to vector spaces.
\end{abstract}


%



\section{Introduction}\label{sec:intr}

\subsection{Combinatorial and algebraic characterizations of entropy}\label{sec:two_faces}
The first part of this paper describes a combinatorial interpretation of the quadratic entropy. It provides an explicit example where the lack of additivity of this function can be explained nonaxiomatically.

 It is well known that Shannon entropy $\ent 1$ is related to the exponential growth of multinomial coefficients. More precisely: given a discrete probability law $(\mu_1,...,\mu_s)$, 
\begin{equation}\label{asymptotics}
\lim_n \frac{1}{n} \ln {n\choose \mu_1n, ...,\mu_sn} = -\sum_{i=1}^s \mu_i \ln \mu_i =: \ent 1(\mu_1,...,\mu_s).
\end{equation}

These coefficients have a $q$-analog. Given an indeterminate $q$, the $q$-integers $\{[n]_q\}_{n\in \Nn}$ are defined by  $[n]_q := (q^n-1)/(q-1)$, the $q$-factorials by $[n]_q! := [n]_q [n-1]_q \cdots [1]_q$, and the $q$-multinomial coefficients are 
\begin{equation}
\qbinom{n}{k_1,...,k_s} := \frac{[n]_q!}{[k_1]_q! \cdots [k_s]_q!},
\end{equation}
where $k_1,...,k_s \in \Nn$ are such that $\sum_{i=1}^s k_i = n$.  When $q$ is a prime power, these coefficients count the number of flags of vector spaces $V_1\subset V_2 \subset ... \subset V_n = \Ff_q^n$ such that $\dim V_i = \sum_{j=1}^i k_j$ (here $\mathbb{F}_q$ denotes the finite field of order $q$); we refer to the sequence $(k_1,...,k_s)$ as the \emph{type} of the flag.  In particular, the $q$-binomial coefficient $\qbinom nk \equiv \qbinom{n}{k,n-k}$ counts vector subspaces of dimension $k$ in $\Ff_q^n$.  

In Section \ref{sec:q_binomials} we study in detail the asymptotic behavior of these coefficients. In particular, we show that, given a discrete probability law $(\mu_1,...,\mu_s)$,  
\begin{equation}\label{a-asymptotics}
\lim_n \frac{2}{n^2} \log_q \qbinom {n}{\mu_1n,...,\mu_sn} = 1- \sum_{i=1}^s \mu_i^2 =:\ent 2(\mu_1,...,\mu_s).
\end{equation}
The function $\ent 2$ is known as \emph{quadratic entropy} \cite{csiszar2008axiomatic}.

More generally, one can introduce a parameterized family of functions $\ln_\alpha: (0,\infty) \to \Rr$, for $\alpha>0$, that generalize the usual logarithm through the formula
\begin{equation}
\ln_{\alpha}(x) := \int_1^x \frac{1}{x^\alpha} \d x.
\end{equation}
The $\alpha$-surprise of a random event of probability $p$ is then defined as $\ln_\alpha(1/p)$, following the traditional definitions in information theory. Given a random variable\footnote{In this work, the range of every random variable is supposed to be a finite set.} $X:\Omega \to S_X$ with law $P$ (a probability on $S_X$), its $\alpha$-entropy $\ent \alpha[X](P)$ is defined as the expected $\alpha$-surprise $\Ee_P\ln_\alpha(1/P(X))$. This $\alpha$-entropy or any real multiple of it can be taken as a generalized information measure. The $1$-entropy is the usual Shannon entropy
\begin{equation}
\ent 1[X](P) = -\sum_{x\in S_X} P(X=x) \ln P(X=x),
\end{equation} 
whereas $\alpha\neq 1$ implies
\begin{equation}
\ent \alpha[X](P)=  \frac{1}{\alpha-1} \left( 1-\sum_{x\in S_X} P(X=x)^\alpha  \right).
\end{equation}
This function appears in the literature under several denominations: it was introduced by Havrda and Charv\'at \cite{havrda} as structural $\alpha$-entropy, Acz\'el and Dar\'oczy \cite{aczel1975measures} call it generalized information function of degree $\alpha$, but by far the most common name is Tsallis $\alpha$-entropy,\footnote{In the physics literature, it is customary to use the letter $q$ instead of $\alpha$, but we reserve $q$ for the 'quantum' parameter that appears in the $q$-integers, $q$-multinomial coefficients, etc.} because Tsallis popularized its use in statistical mechanics. 

Given a second variable $Y:\Omega \to S_Y$ and a law $P$ for the pair $(X,Y)$, the $\alpha$-entropy satisfy the equations
\begin{equation}\label{cocycle_eqns}
\begin{split}
\ent \alpha[(X,Y)](P) { }={ }& \ent \alpha[X](X_* P) \\
& + \sum_{x\in S_X} P(x)^\alpha \ent \alpha[Y](Y_* P|_{X=x}) 
\end{split}
\end{equation}
where $P(x) := P(\{X=x\})$, the symbol $P|_{X=x}$ denotes the conditional law, and $X_*Q$ is the push-forward of the law $Q$ on $S_X\times S_Y$ under the canonical projection $\pi_X :S_X\times S_Y \to S_X$. We have shown in \cite{vigneaux2017generalized} that $\ent \alpha[\cdot]$ is the only family of measurable real-valued functions that satisfy these functional equations for generic collections of random variables and probabilities, up to a multiplicative constant $K$. The case $\alpha = 1$ is already treated in \cite{bennequin}. Of course, this depends on a long history of axiomatic characterizations of entropy that begins with Shannon himself, see \cite{shannon1948,aczel1975measures,csiszar2008axiomatic}.

In particular, if $X$, $Y$ represent the possibles states of two independent systems (e.g. physical systems,  random sources), in the sense that $P(x,y) = X_*P(x) Y_*P(y)$, then
\begin{equation}\label{additivity}
\ent 1[(X,Y)](P) = \ent 1[X](X_*P)+\ent 1[Y](Y_*P).
\end{equation}
This property of Shannon entropy is called additivity. Under the same assumptions, Tsallis entropy verifies (for $K=1$):
\begin{equation}\label{non-additivity}
\begin{split}
\ent \alpha[(X,Y)](P) { }={ }& \ent \alpha[X](X_*P)+\ent \alpha[Y](Y_*P) \\ 
&- (\alpha-1) \ent \alpha[X](X_*P)\ent \alpha[Y](Y_*P).
\end{split}
\end{equation}
It is said that Tsallis entropy is nonadditive.\footnote{Originally, this was called \emph{non-extensivity}, which explains the name `nonextensive statistical mechanics'.} This property is problematic from the point of view of heuristic justifications for information functions, that have always assumed as `intuitive' that the amount of information given by two independent events should be computed as the sum of the amounts of information given by each one separately (this explains the use of the logarithm to define the surprise).

The initial motivation behind this paper was to understand better these generalized information functions of degree $\alpha$. Tsallis  used them as the foundation of nonextensive statistical mechanics, a generalization of Boltzmann-Gibbs statistical mechanics that was expected to describe well some systems with long-range correlations. It is not completely clear which kind of statistical systems follow these ``generalized statistics''.\footnote{``...the entropy to be used for thermostatistical purposes would be not universal but would depend on the system or, more precisely, on the nonadditive universality class to which the system belongs.”\cite[p. xii]{tsallis-book}}  There is extensive empirical evidence about the pertinence of the predictions made by nonextensive statistical mechanics \cite{tsallis-book}. However, very few papers address the microscopical foundations of the theory (for instance, \cite{ruiz2015emergence,hanel2011generalized,jensen2016statistical}). We present here a novel approach in this direction, based on the combinatorics of flags, but only for the case $\alpha = 2$. However, we indicate in the last section how these ideas could be extended to other cases.

There is a connection between the combinatorial and algebraic characterizations of entropy, that we describe in Section \ref{sec:combinatorial_shannon} (Shannon entropy) and Section \ref{sec:combinatorial_explanation} (quadratic entropy).  The well known multiplicative relations at the level of multinomial coefficients shed new light on additivity/nonadditivity. In the simplest case, let $(p_0,p_1)$, $(q_0,q_1)$ be two probability laws on $\{0,1\}$; then
\begin{equation}\label{multiplicative_relations}
{n \choose p_0q_0 n,p_0q_1 n,p_1q_0n,p_1q_1n} = {n \choose p_0n} {p_0n \choose p_0q_0n }{p_1n \choose p_1q_0n}.
\end{equation}
Applying $\frac 1n \ln (-)$ to both sides and taking the limit $n\to \infty$, we recover \eqref{additivity}. Equation \eqref{multiplicative_relations} remains valid for the $q$-multinomial coefficients, but in this case one should apply $\lim_n \frac{2}{n^2} \log_q(-)$ to both sides to obtain the quadratic entropy:
\begin{align*}\label{non-add-binary}
\ent 2 & (p_0q_0,p_0q_1,p_1q_0,p_1q_1)  \\&= \ent 2(p_0,p_1) + p_0^2 \ent 2(q_0,q_1) + (1-p_0)^2 \ent 2(q_0,q_1) \\
&= \ent 2(p_0,p_1) + \ent 2(q_0,q_1) - \ent 2(p_0,p_1)\ent 2(q_0,q_1).
\end{align*}
Thus, asymptotically, the number of flags $V_{00}\subset V_{01} \subset V_{10}\subset V_{11} = \Ff_q^n$ of type $(p_0q_0 n,p_0q_1 n,p_1q_0n,p_1q_1n)$ can be computed in terms of the number of flags $W_0\subset W_1=\Ff^n_q$ of type $(p_0n,p_1n)$ and those flags $W'_0 \subset W_1' = \Ff_q^m$ of type $(q_0m,q_1m)$ |where $m$ can take the values $p_0n$ or $p_1n$| through this nonadditive formula. 

\subsection{A $q$-deformation of Shannon's theory} 
The second part of this paper builds a generalization of Shannon's theory where messages are vector spaces. Formula \eqref{a-asymptotics} already suggests that the quadratic entropy plays an essential role in it.

In fact, the corresponding formula \eqref{asymptotics} is of great importance in Shannon's theory. Consider a random source that emits at time $n\in \Nn$ a symbol $Z_n$ in $S_Z=\{z_1,...,z_s\}$, each $Z_n$ being an independent realization of a $S_Z$-valued random variable $Z$ with law $P$. A \emph{message} (at time $n$) corresponds to a random sequence $(Z_1,...,Z_n)$ taking values in $S_Z^n$ with law $P^{\otimes n}$. The \emph{type} of a sequence $\v z\in S_Z^n$ is the probability distribution on $S_Z$ given by the relative frequency of appearance of each symbol in it; for example, when $S_Z = \{0,1\}$, the type of a sequence with $k$ ones is $(1-\frac k n) \delta_0 +  \frac k n\delta_1$. A ``typical sequence''  is expected to have type $P$, and therefore its probability $P^{\otimes n}(\v z)$ is approximately $\prod_{z\in S_Z} P(z)^{nP(z)} = \exp\{-n \ent 1[Z](P)\}$.  The cardinality of the set of sequences of type $P$ is ${n \choose P(z_1)n,...,P(z_s)n} \approx \exp\{n \ent 1[Z](P)\}$. This implies, according to Shannon, that ``it is possible for most purposes to treat the long sequences as though there were just $2^{Hn}$ of them, each with a probability $2^{-Hn}$" \cite[p. 397]{shannon1948}. This result is known nowadays as the asymptotic equipartition property (AEP), and can be stated more precisely as follows   \cite[Th. 3.1.2]{cover2006elements}: given $\epsilon>0$ and $\delta>0$, it is possible to find $n_0 \in \Nn$ and sets $\{A_n\}_{n\geq n_0}$, $A_n \subset S_Z^n$, such that, for every $n\geq n_0$,
\begin{enumerate}
\item  $P^{\otimes n}(A_n^c) < \epsilon$, and 
\item for every $\v z \in A_n$,  
\begin{equation}\label{eq:class_AEP_equiprobability}
\left|\frac 1n \ln(P^{\otimes n}(\v z)) - \ent 1[Z](P)\right| < \delta.
\end{equation}
\end{enumerate} 
Furthermore, if $s(n,\epsilon)$ denotes
\begin{displaymath}
 \min \set{|B_n| }{B_n\subset S_Z^n \text{ and } \Pr{(Z_1,...,Z_n) \in B_n} \geq 1-\epsilon},
\end{displaymath} 
then
\begin{equation}
 \lim_n \frac{1}{n} \ln |A_n| = \lim_n \frac{1}{n} \ln s(n,\epsilon) =  \ent 1[Z](P).
\end{equation}
The set $A_n$ can be defined to contain all the sequences whose type $Q$ is close to $P$, in the sense that $\sum_{z\in S_Z} |Q(z)-P(z)|$ is upper-bounded by a small quantity; this is known as \emph{strong typicality} (see \cite[Def. 2.8]{csiszar1981information}).

Similar conclusions can be drawn for a system of $n$ independent physical particles, the state of each one being represented by a random variable $Z_i$; in this case, the vector $(Z_1,...,Z_n)$ is called a \emph{configuration}. The set $A_n$ can be thought as an approximation to the effective phase space (``reasonable probable'' configurations) and the entropy as a measure of its size, see \cite[Sec. V]{jaynes1965gibbs}.  In both cases |messages and configurations| the underlying probabilistic model is a process  $(Z_1,...,Z_n)$ linked to the multinomial distribution, and the AEP is merely a result on measure concentration around the expected type. 

We envisage a new type of statistical model, such that a message at time $n$ (or a configuration of $n$ particles) is represented by  a flag of vector spaces $V_1 \subset V_2 \subset ... \subset V_s = \Ff_q^n$. In the simplest case ($s=2$) a message is just a vector space $V$ in $\Ff_q^n$. While the type of a sequence is determined by the number of appearances of each symbol, the type of a flag is determined by its dimensions or |equivalently| by the numbers $(k_1,...,k_s)$ associated to it; by abuse of language, we refer to $(k_1,...,k_s)$ as the type. The cardinality of the set of flags $V_1 \subset...\subset V_s \subset\Ff_q^n$ that have type $(k_1,...,k_s)$ is  $\qbinom n{k_1,...,k_s} \sim C(q) q^{n^2 \ent 2(k_1/n,...,k_s/n)/2}$, where $C(q)$ is an appropriate constant. 

To build a correlative of Shannon's theory of communication, it is fundamental to have a probabilistic model for the source. In our case, this means a random process $\{F_i\}_{i\in \Nn}$ that produces at time $n$ a flag $F_n$ that would correspond to a generalized message. We can define such process if we restrict our attention to the binomial case ($s=2$). This is the purpose of Section \ref{sec:dynamical_model}.

Let $\theta$ be a positive real number, and let $\{X_i\}_{i\geq 1}$ be a collection of independent random variables that satisfy $X_i\sim \Ber\left(\frac{\theta q^{i-1}}{1+\theta q^{i-1}}\right)$, for each $i$. We fix a a sequence of linear embeddings $\Ff_q^1 \hookrightarrow \Ff_q^2 \hookrightarrow ...$, and identify $\Ff_q^{n-1}$ with its image in $\Ff_q^n$. We define then a stochastic process $\{V_i\}_{i\geq 0}$ such that each $V_i$ is a vector subspace of $\Ff_q^i$, as follows: $V_0 = 0$ and, at step $n$, the dimension of $V_{n-1}$ increases by $1$ if and only if $X_n = 1$; in this case, $V_n$ is picked at random (uniformly) between all the $n$-dilations of $V_{n-1}$. When $X_n=0$, one sets $V_n = V_{n-1}$. The $n$-dilations of a subspace $w$ of $\Ff_q^{n-1}$  are defined as
\begin{equation}
\begin{split}
\Dil_n (w) = \set{v\subset \Ff_q^n}{ & \dim v - \dim w = 1, \\  & w\subset v\text{ and  } v \not \subset \Ff_q^{n-1}}.
\end{split}
\end{equation} 
We prove that, for any subspace $v\subset \Ff_q^n$ of dimension $k$,  $\Pr{V_n = v}=\frac{\theta^k q^{k(k-1)/2}}{(-\theta;q)_n}$. This implies that $\Pr{\dim V_n =k} =\qbinom nk \frac{\theta^k q^{k(k-1)/2}}{(-\theta;q)_n}$, which appears in the literature as $q$-binomial distribution. (We have used here the $q$-Pochhammer symbols $(a;q)_n := \prod_{i=0}^{n-1} (1-aq^i)$, with $(a;q)_0  = 1$.)

\begin{table}[!t]
\caption{Correspondence between Shannon's information theory in the case of memoryless Bernoulli sources and our $q$-deformed version for vector spaces.}
\label{table_correspondence}
\centering
\renewcommand*{\arraystretch}{1.5}
\renewcommand{\tabularxcolumn}[1]{m{#1}}
\newcolumntype{C}{>{\centering\arraybackslash}X}
\begin{tabularx}{\columnwidth}{|C|C|C|}
\hline \textbf{Concept} & \textbf{Shannon case} & \textbf{$q$-case} \\\hline 
Message at time $n$ ($n$-message) & Word $w\in\{0,1\}^n$ & Vector subspace $v\subset F_q^n$ \\ \hline
Type    & Number of ones & Dimension \\ \hline
Number of $n$-messages of type $k$ & $\displaystyle{n\choose k} $ & $\displaystyle\qbinom{n}{k}$ \\ \hline
Probability of a $n$-message of type $k$ & $\displaystyle \xi^k (1-\xi)^{n-k}$ & $\displaystyle\frac{\theta^k q^{k(k-1)/2}}{(-\theta;q)_n}$ \\ \hline
\end{tabularx}
\end{table}

For the multinomial process, the probability $P^{\otimes n}$ concentrates on types close to $P$ i.e. appearances close to the expected value $nP(z)$, for each $z\in S_Z$. In the case of $V_n$, the probability also concentrates on a restricted number of dimensions (types). In fact, it is possible to prove an analog of the asymptotic equipartition partition property; this is the main result of this work, Theorem \ref{them:AEP}. It can be paraphrased as follows: 

 \textit{for every $\delta>0$ and almost every $\epsilon>0$ (except a countable set), there exist $n_0\in \Nn$ and sets $A_n = \bigcup_{k=0}^{ \Delta(p_\epsilon)} \Gr(n-k,n)$, for all $n\geq n_0$, such that $\Delta(p_\epsilon)$ is an integer that just depends on $\epsilon$, $\Pr{V_n\in A_n^c} \leq \epsilon$ and, for any $v\in A_n$ such that $\dim v = k$,
\begin{equation}
\left| \frac{\log_q (\Pr{V_n = v}^{-1})}{n} - \frac{n}{2}\ent 2(k/n) \right| \leq \delta.
\end{equation} 
Moreover, the size of $A_n$ is optimal, up to the first order in the exponential: let $s(n,\epsilon)$ denote 
$$\min\set{|B_n| }{B_n\subset \Gr(n) \text{ and } \Pr{V_n \in B_n} \geq 1-\epsilon},$$  then
\begin{align*}
\lim_n \frac{1}{n} \log_q|A_n| &{ }={ } \lim_n \frac{1}{n} \log_q s(n,\epsilon)  \\ 
&{ }={ } \lim_n \frac{n}{2} \ent 2(\Delta(p_\epsilon)/n) \\ 
&{ }={ } \Delta(p_\epsilon). 
\end{align*}}

The set $A_n$ correspond to the ``typical subspaces'', in analogy with the typical sequences introduced above. We close Section \ref{sec:generalized_info} with an application of this theorem to source coding.

\subsection{Notation}

The statement  $A:=B$ means that $A$ is defined to be $B$, as well as $B=:A$. For us $\Nn=\{0,1,2,3,...\}$. The symbols $\Rr$, $\Cc$ and $\Ff_q$ denote respectively the field real numbers, the field of complex numbers, and the Galois field of order $q$. The multinomial coefficients are denoted by ${n \choose k_1,...,k_s}$, and ${n\choose k} := {n \choose k, n-k}$ are the binomial coefficients. The $q$-multinomial coefficients $\qbinom{n}{k_1,...,k_s}$ are defined in Section \ref{subsec:q_binomials_def}, together with the $q$-binomial coefficients $\qbinom{n}{k}$. The discrete interval $\llbracket a, b\rrbracket$ corresponds to $\Zz \cap [a,b]$ as a subset of $\Rr$. We use Iverson's convention for characteristic functions \cite{knuth1992two}: for any proposition $p$,
$$[p]:=\begin{cases} 1 & \text{if } p\text{ is true}\\ 0 & \text{if } p\text{ is false} \end{cases}.$$
Therefore the indicator functor of a set $B$ corresponds to $x\mapsto [x\in B]$. The symbol $\log_q$ signifies the usual logarithm in base $q$, as opposed to Tsallis' deformed $\alpha$-logarithm $\ln_\alpha$. Finally, $f_n\sim g_n$ means $f_n/g_n\to 1$ as $n\to \infty$


\section{Combinatorial characterization of Shannon's information}\label{sec:combinatorial_shannon}

Let  $X$ be a finite random variable that takes values in the set $S_X=\{x_1,...,x_s\}$. We suppose that, among $N$ independent trials of the variable $X$, the result $x_i$ appears $N(x_i)$ times, for each $i$. Evidently, $\sum_i N(x_i)=N$. 

The number of sequences in $(S_X)^N$ that agree with the prescribed counting $(N(x_1),...,N(x_s))$ is given by the multinomial coefficient
\begin{equation}\label{multinomial}
{N \choose \{N(x_k)\}_{k=1}^s } := \frac{N!}{N(x_1)! \cdots N(x_s)!}.
\end{equation}

But we could also reason iteratively. Let us consider a partition of $\{x_1,...,x_s\}$ in $t$ disjoint sets, denoted $Y_1,..., Y_t$. These can be seen as level sets of a new variable $Y$, taking values in a set $S_Y=\{y_1,...,y_t\}$; by definition, $\{Y=y_t\}=Y_t$. There is surjection $\pi:S_X\to S_Y$ that sends $x\in S_X$ to the unique  $y\in S_Y$ such that $x\in \{Y=y\}$. The probability $\nu(x_i) = N(x_i)/N$ on $S_X$ can be pushed-forward under this surjection; the resulting law $\pi_* \nu$ satisfies $\pi_*\nu (y) = \sum_{x\in \pi^{-1}(y)} \nu(x)$. Our counting problem can be solved as follows: count first the number of sequences in $(S_X)^N$ such that $N\pi_*\nu(y_i)$ values correspond to the group $y_i$, for $i\in \{1,...,t\}$. This equals
\begin{equation}
{N \choose \{N\pi_*\nu(y_i)\}_{i=1}^t }.
\end{equation}
Then, for each group $\pi^{-1}(y_i)\equiv \{Y=y_i\}$, count the number of sequences of length $N\pi_*\nu(y_i)$ (subsequences of the original ones of length $N$) such that every $x_j \in \pi^{-1}(y_i)$ appears $N(x_j)$ times. These are 
\begin{equation}
{N\pi_*\nu(y_i) \choose \{N(x_j)\}_{x_j \in \pi^{-1}(y_i)} }.
\end{equation} 
In total, the number of  sequences of length $N$ such that $x_k$ appears $N(x_k)$ times, for every $k\in \{0,...,s\}$, are
\begin{equation}\label{discrete_cocyle}
{N \choose \{N\pi_*\nu(y_i)\}_{i=1}^t } \prod_{i=1}^t {N\pi_*\nu(y_i) \choose \{N(x_j)\}_{x_j \in \pi^{-1}(y_i)} }.
\end{equation}
The considerations above give the identity
\begin{multline}\label{multiplicative_identity_multinomials}
{N \choose \{N(x_k)\}_{k=1}^s } = \\ {N \choose \{N\pi_*\nu(y_i)\}_{i=1}^t } \prod_{i=1}^t {N\pi_*\nu(y_i) \choose \{N(x_j)\}_{x_j \in \pi^{-1}(y_i)} }.
\end{multline}
This can be rephrased as follows: the multinomial expansion of $(x_1 + \cdots + x_s)^N$ and the iterated multinomial expansion of $( \sum_{y_i} ( \sum_{x_j\in \pi^{-1}(y_i)} x_j ))^N$ assign the same coefficient to $x_1^{N(x_1)}x_2^{N(x_2)}\cdots x_s^{N(x_s)}$.

Equation \eqref{multiplicative_identity_multinomials} implies that
\begin{equation}\label{additive_relation_preH}
\begin{split}
\frac{1}{N} &\log {N \choose \{N(x_k)\}_{k=1}^s } { }={ } \frac{1}{N}\log {N \choose \{N\pi_*\nu(y_i)\}_{i=1}^t }  \\
 & + \sum_{i=1}^t \pi_*\nu(y_i)\frac{1}{N\pi_*\nu(y_i)} \log {N\pi_*\nu(y_i) \choose \{N(x_j)\}_{x_j \in \pi^{-1}(y_i)} }. 
 \end{split}
\end{equation}
\normalsize
We can see this as a discrete analog of the third axiom of Shannon. The connection is made explicit by means of the following proposition.

\begin{proposition}\label{log-asymptotics-H1}
Let $N$ be a natural number and $\{N(i)\}_{i=0}^s$ such that $\sum_{i=0}^s N(i)= N$. Suppose that $N(i)/N \to \mu_i \in [0,1]$ as $N\to \infty$, for all $i$. Then
\begin{equation}
\lim_{N\to \infty} \frac{1}{N} \ln {N \choose N(0), ...., N(s) } = \ent 1(\mu_0,\cdots, \mu_s),
\end{equation}
where $\ent 1$ denotes Shannon entropy:
\begin{equation}
\ent 1(\mu_0,\cdots, \mu_s) = -\sum_{i=0}^{s} \mu_i \ln \mu_i.
\end{equation}
\end{proposition}
\begin{IEEEproof}
This is a standard result. See for example \cite[Lemma 4.1]{mori}.
\end{IEEEproof}

By convention, $0\ln 0 = 0$. If we take the limit of \eqref{additive_relation_preH} under the hypotheses of the previous proposition, we obtain
\begin{multline}\label{functional_equations}
\ent 1(\mu(x_1),...,\mu(x_s)) = \ent 1(\pi_*\mu(y_1),...,\pi_*\mu(y_t)) \\ +  \sum_{i=0}^t \pi_*\mu(y_i) \ent 1(\mu|_{Y=y_i}(x_1),...,\mu|_{Y=y_i}(x_s)).
\end{multline}

Consider now the particular case $X=(Z,Y)$, for certain random variable $Z$ taking values on $S_Z$. We use the notations introduced in  Section \ref{sec:two_faces}. Since the support of $\mu|_{Y=y_i}$ is $S_Z\times \{y_i\}$, isomorphic to $S_Z$ by the natural projection $\pi_Z: S_Z\times S_Y \to S_Z$, there is a clear identification of $\ent 1(\mu|_{Y=y_i}(x_1),...,\mu|_{Y=y_i}(x_s))$ with $H[Z](Z_* \mu|_{Y=y_i})$. Therefore,  \eqref{functional_equations} reads
\begin{equation}\label{functional_equations_fancy}
\ent 1[(Z,Y)](\mu) = \ent 1[Y](Y_*\mu) +  \sum_{i=0}^t Y_*\mu(y_i) \ent 1[Z](Z_* \mu|_{Y=y_i}).
\end{equation}
This proves combinatorially that Shannon entropy satisfy all the functional equations of the form \eqref{functional_equations}. The ensemble of these equations |for a given family of finite sets and surjections between them| constitute a cocycle condition in information cohomology (see \cite{bennequin} and \cite{vigneaux2017generalized}). These are functional equations that have as unique solution Shannon entropy. 

In the following sections, we explore a generalization of the previous argument. We replace the multinomial coefficient with their $q$-deformation, that satisfy the same multiplicative relations. These $q$-multinomial coefficients are related asymptotically to the quadratic entropy. 

\section{The $q$-multinomial coefficients}\label{sec:q_binomials}

We introduce here the combinatorial objects and results used throughout this article. In Section \ref{subsec:q_binomials_def}, we define the $q$-multinomial coefficients, that are associated to the enumeration of flags of finite vector spaces. Section \ref{sec:asymptotic_behavior} studies their asymptotic behavior and establishes the connection with the quadratic entropy. Sections \ref{sec:combinatorial_explanation} and \ref{sec:max_ent} are mutually independent and not essential to understand the rest of the paper: the former  uses the asymptotic results to obtain a combinatorial explanation for the nonadditivity of Tsallis $2$-entropy, and the later discuss a combinatorial justification of the maximum entropy principle with Tsallis entropy.

\subsection{Definition}\label{subsec:q_binomials_def} Let $q$ be an indeterminate. Given $(n, k_1,...,k_s)\in  \Nn^{s+1}$ such that $\sum_{i=1}^s k_i = n$, the $q$-multinomial coefficient ${\qbinom n {k_1,...,k_s}}$ is defined by the formula
\begin{equation}
\qbinom n {k_1,...,k_s} := \frac{[n]_q!}{[k_1]_q! \cdots [k_s]_q!}.
\end{equation}
We have used the notation for $q$-factorials introduced in Section \ref{sec:intr}. 

Throughout this paper, we shall assume that $q$ is a fixed prime power. For such $q$, the $q$-binomial coefficient $\qbinom nk \equiv \qbinom n{k,n-k}$ counts the number of $k$-dimensional subspaces in $\Ff_q^n$. More generally, given a set of integers $k_1,...,k_s$ such that $\sum_{i=1}^s k_i = n$, the $q$-multinomial coefficient $\qbinom{n}{k_1,...,k_s}$ equals the number of flags $V_1\subset V_2 \subset \cdots \subset V_{s-1} \subset V_s = \Ff_q^n$ of vector spaces such that $\dim V_j = \sum_{i=1}^j k_i$ \cite{Prasad2010-1, Prasad2010-2}. We will say that these flags are of \textit{type } $(k_1,...,k_s)$. 

It is possible to introduce a function $\Gamma_q$ as the normalized solution of a functional equation that guaranties that $[n]_q ! = \Gamma_q(n+1)$, see \cite{askey}. When $q>1$  and $x>0$, this function is given by the formula \cite{moak}:
\begin{align}\label{qGamma}
\Gamma_q(x) &= (q^{-1};q^{-1})_\infty q^{{x\choose 2}} (q-1)^{1-x} \sum_{n=0}^\infty \frac{q^{-nx}}{(q^{-1};q^{-1})_n} \\ &= \frac{(q^{-1};q^{-1})_\infty q^{{x\choose 2}} (q-1)^{1-x}}{(q^{-x};q^{-1})_\infty} ,
\end{align}
where we have used the Pochhammer symbol
\begin{equation}
(a;x)_n := \prod_{k=0}^{n-1} (1-ax^k), \qquad (a;x)_0 = 1.
\end{equation}
The equivalent expressions for the  function $\Gamma_q$ come from the identity
\begin{equation}\label{eq:q-binomial-theorem}
\frac{(ax;q)_\infty}{(x;q)_\infty} =\sum_{n=0}^\infty \frac{(a;q)_n}{(q;q)_n}x^n \qquad(|q|<1),
\end{equation}
known as $q$-binomial theorem (see 
\cite[p. 30]{kac2001quantum}).

Recall \cite[p. 92]{knopp} that an infinite product $\prod_{i=0}^\infty u_i$ is said to be convergent if
\begin{enumerate}
\item there exists $i_0$ such that $u_i\neq 0$ for all $i>i_0$;
\item $\lim_{n\to \infty} u_{i_0+1} \cdots u_{i_0 +n}$ exists and is different from zero.
\end{enumerate}
An infinite product in the form $\prod (1+c_i)$ is said to be absolutely convergent when $\prod (1+|c_i|)$ converges. One can show that absolute convergence implies convergence. Moreover, when the terms $\gamma_i \geq 0$, the product $\prod_i (1 + \gamma_i)$ is convergent if and only if the series $\sum_i \gamma_i$ converges. The convergence of $\sum_i 1/q^i$ gives then the following result, that is used without further comment throughout the paper.

\begin{lemma}
For every $a\in \Cc$, the product $(a;q^{-1})_\infty$ converges. Moreover, if $a\not\in \set{q^i}{i\geq 0}$, then $(a;q^{-1})_\infty \neq 0$.
\end{lemma}

The $\Gamma_q$ function gives an alternative expression for the $q$-multinomial coefficients
\begin{equation}\label{q-coeff-qgamma}
\qbinom{n}{k_1,...,k_s} = \frac{\Gamma_q(n+1)}{\Gamma_q(k_1+1) \cdots \Gamma_q(k_s+1)},
\end{equation}
which in turn extends its definition to complex arguments.

We close this subsection with a remark on the unimodality of the $q$-binomial coefficients.
\begin{lemma}\label{lemma:unimodality_binomial}
For every $n\in \Nn$,
\begin{itemize}
\item $\qbinom n0 < \qbinom n1 < \ldots <  \qbinom n{\floor{n/2}},$
\item $ \qbinom n{\floor{n/2}}  = \qbinom n{\ceil{n/2}}, $
\item $ \qbinom n{\ceil{n/2}} > \ldots \qbinom n {n-1} > \qbinom nn.$
\end{itemize}
\end{lemma}
\begin{IEEEproof}
Consider the quotient 
\begin{equation}
q(n,k):= \frac{\qbinom n{k+1}}{\qbinom nk} = \frac{[n-k]_q}{[k+1]_q}.
\end{equation}
Then, $q(n,k)\geq 1$ iff $q^{n-k}\geq q^{k+1}$ iff $k \leq \frac{n-1}{2}$, with equality just in the case $k=\frac{n}{2}-\frac 12=\floor{n/2}$ (when $n$ is odd).
\end{IEEEproof}
\subsection{Asymptotic behavior}\label{sec:asymptotic_behavior}

The quadratic entropy $\ent 2$ of a probability law $(\mu_1,...,\mu_s)$ is defined by the formula: \footnote{We fix the constant $1$ in front of $1- \sum_{i=1}^s \mu_i^2$. In \cite{vigneaux2017generalized} we have characterized Tsallis $\alpha$-entropy ($\alpha>0$) with system of functional equations (as a $1$-cocycle in cohomology), whose general solution is $\frac{K}{2^{1-\alpha}-1} \left(1- \sum_{i=1}^s \mu_i^\alpha\right)$, for $K$ an arbitrary constant.}
\begin{equation}
\ent 2(\mu_1,\cdots, \mu_s) := 1- \sum_{i=1}^s \mu_i^2.
\end{equation}

\begin{theorem}\label{prop:asymptotic_behavior}
For each $n\in \Nn$, let $\{k_i(n)\}_{i=1}^s $ be a set of positive real numbers such that $\sum_{i=0}^s k_i= n$ (we write $k_i$ when $n$ is clear from context). Suppose that, for each $i\in\{1,...,s\}$, it is verified that $k_i(n)\to l_i \in [0,\infty]$ as $n\to \infty$. Then,
\begin{multline*}
\qbinom{n}{k_1,...,k_s} \sim  \\(q^{-1};q^{-1})_\infty^{1- s}\prod_{i =1}^s  (q^{-(l_i +1)};q^{-1})_\infty  q^{n^2 \ent 2(\frac {k_1}{n},...,\frac{k_s}{n})/2}.
\end{multline*}
\end{theorem}
Recall that $f_n\sim g_n$ means $f_n/g_n\to 1$ as $n\to \infty$. By convention, $(q^{-(\infty +1)};q^{-1})_\infty = 1$.
\begin{IEEEproof} See Appendix \ref{proof_prop_asymptotic_q_multinomial}.
\end{IEEEproof}

When $f_n$ and $g_n$ are positive, $f_n \sim g_n$ implies that $\lim_n \frac{1}{n}(\log_q f_n - \log_q g_n) = 0$. For instance, we can deduce  that, for any fixed $\Delta \in \Nn$, 
\begin{equation}\label{computation_symptotics_binomial}
\lim_n \frac 1n \log_q \qbinom{n}{n-\Delta} = \lim_n  \frac n 2 \ent 2(\Delta/n) = \Delta,
\end{equation}
where the last equality comes from a direct computation. 

As an immediate application of Theorem \ref{prop:asymptotic_behavior}, we obtain the equivalent of Proposition \ref{log-asymptotics-H1}.

\begin{proposition}
For each $n\in \Nn$, let $\{k_i(n)\}_{i=1}^s $ be a set of positive real numbers such that $\sum_{i=0}^s k_i= n$ (we write $k_i$ when $n$ is clear from context). Suppose that $k_i/n \to \mu_i \in [0,1]$ as $n\to \infty$, for all $i$. Then
\begin{equation}
\lim_{n\to \infty}  \frac{2}{n^2} \log_q \qbinom{n}{k_1, ...., k_s } =  \ent 2(\mu_1,\cdots, \mu_s).
\end{equation}
\end{proposition}
\begin{IEEEproof}
If $f/g\to 1$, then $\log_q(f/g) \to 0$. Therefore, 
\begin{multline}
\log_q\qbinom{n}{k_1,...,k_s} - \log_q\left(\frac{(q^{-1};q^{-1})_\infty^{1- s}}{\prod_{i =1}^s  (q^{-(l_i +1)};q^{-1})_\infty}\right) \\ -   \frac{n^2}{2} \ent 2\left(\frac{k_1}{n},...,\frac{k_s}{n}\right) = o(1).
\end{multline}
Multiply this by $2/n^2$ and use the continuity of $\ent 2$ to conclude. 
\end{IEEEproof}

\subsection{Combinatorial explanation for nonadditivity of Tsallis 2-entropy}\label{sec:combinatorial_explanation}

Additivity corresponds to the following property of Shannon entropy: if $X$ is a random variable with law $P=\{p_x\}_{x\in S_X}$ and $Y$ another with law $Q=\{q_y\}_{y\in S_Y}$, independent of  $X$, then the joint variable $(X,Y)$ has law $P\otimes Q=\{p_xq_y\}_{(x,y)\in S_X\times S_Y}$ and 
\begin{equation}\label{additivity_H1}
\ent 1[(X,Y)](P\otimes Q) = \ent 1[X](P) + \ent 1[Y](Q).
\end{equation}

For simplicity (the arguments work in general), we suppose that  $X$, $Y$ are binary variables, i.e. $S_X=S_Y = \{0,1\}$. Consider the sequences counted by ${N\choose N_{00}, N_{01}, N_{10}, N_{11}}$; they are the possible results of $N$ independent trials of the variable $(X,Y)$, under the assumption that the result $(i,j)$ is obtained $N_{ij}$ times, for each $(i,j)\in \{0,1\}^2$. We treat the particular case $N_{ij} = p_iq_jN$, that correspond to the expected number of appearances of $(i,j)$. The independence between $Y$ and $X$ means that, given $N_0:=N_{00}+N_{01} = p_0 N$ occurrences of $X=0$ (resp. $N_1:=N_{10}+N_{11}=p_1N$ occurrences of $X=1$) in the sequences of length $N$ counted above, there are $q_0N_i$  occurrences of $Y=0$ and $q_1N_i$ occurrences of $Y=1$ in the corresponding subsequence defined by the condition $X=i$, irrespective of the value of $i$. In this case, \eqref{multiplicative_identity_multinomials} specializes to 
\begin{equation}\label{iterative_multinomial}
{N\choose N_{00}, N_{01}, N_{10}, N_{11}} = {N \choose N_0} {N_0 \choose q_0 N_0 }{N_1 \choose q_0N_1}.
\end{equation}
Applying $\frac 1N \ln (-)$ to both sides and taking the limit $N\to \infty$, we recover  \eqref{additivity_H1}. (This is just a particular case of the computations in Section \ref{sec:combinatorial_shannon}.)

In the $q$-case, $\qbinom{N}{N_{00}, N_{01}, N_{10}, N_{11}}$ counts the number of flags $V_{00}\subset V_{01}\subset V_{10}\subset V_{11} = \Ff_q^n$ of type $(N_{00}, N_{01}, N_{10}, N_{11})$. When $N_{ij} = p_iq_j N$, such a flag can be determined by an iterated choice of subspaces, whose dimensions are chosen independently:   pick first a subspace $V_0 \subset \Ff_q^n$ of dimension $N_0=N_{00}+N_{01} = p_0 N$ (there are $\qbinom{N}{N_0}$ of those) and then pick a subspace of dimension $ q_0N_0 \subset V_0$ and another subspace of dimension $q_0N_1$ in $\Ff_q^n/V_0$. This corresponds to the combinatorial identity
\begin{equation}\label{iterative_qmultinomial}
\qbinom{N}{N_{00}, N_{01}, N_{10}, N_{11}} = \qbinom {N}{N_0} \qbinom {N_0}{q_0 N_0} \qbinom {N_1}{q_0 N_1}.
\end{equation}
Applying $\frac 2{N^2} \log_q (-)$ to both sides and taking the limit $N\to \infty$, we obtain
\begin{align*}
\ent 2 & (p_0q_0,p_0q_1,p_1q_0,p_1q_1)  \\&= \ent 2(p_0,p_1) + p_0^2 \ent 2(q_0,q_1) + (1-p_0)^2 \ent 2(q_0,q_1) \\
&= \ent 2(p_0,p_1) + \ent 2(q_0,q_1) - \ent 2(p_0,p_1)\ent 2(q_0,q_1).
\end{align*}
In both cases, the trees that represent the iterated counting are the same, see Fig. \ref{fig:tree} (and compare this with Figure 6 in Shannon's paper \cite{shannon1948}). The main difference lies in the exponential growth of the combinatorial quantity of interest and how the correspondent exponents are combined. In the $q$-case, even if you choose the dimensions in two independent steps, the exponents do not simply add; in fact, the counting of sequences is nongeneric in this respect. Remark also that the interpretation of probabilities as relative \emph{frequencies} of symbols only make sense for the case of words; more generally they correspond to ratios or relative proportions.

\tikzstyle{level 1}=[level distance=4cm, sibling distance=2cm]
\tikzstyle{level 2}=[level distance=3cm, sibling distance=1.5cm]

\tikzstyle{end} = [circle, minimum width=3pt,fill, inner sep=1.2pt]

\begin{figure}
\centering
\begin{tikzpicture}[grow=right, sloped]
\node[end] {}
    child {
        node[end] {}        
            child {
                node[end] {}
                edge from parent
                node[below] {$q_0$}
            }
            child {
                node[end] {}
                edge from parent
                node[above]  {$q_1$}
            }
            edge from parent 
            node[below] {$p_0$}
    }
    child {
        node[end] {}        
        child {
                node[end] {}
                edge from parent
                node[below] {$q_0$}
            }
            child {
                node[end] {}
                edge from parent
                node[above]  {$q_1$}
                }
        edge from parent         
            node[above]  {$p_1$}
    };
\end{tikzpicture}
\caption{Decision tree for the recursive reasoning that leads to equations \eqref{iterative_multinomial} and \eqref{iterative_qmultinomial}.}\label{fig:tree}
\end{figure}
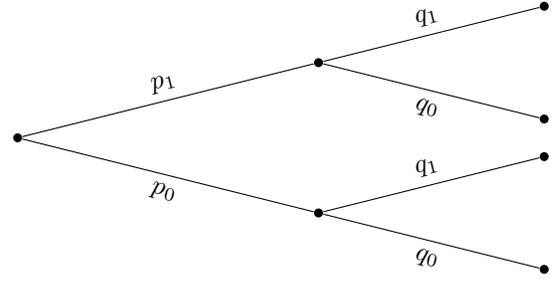


%
%

\subsection{Maximum entropy principle}\label{sec:max_ent}

In the simplest models of statistical mechanics, one assumes that the system is composed of $n$ particles,  each one in certain state from a finite set $S=\{s_1,...,s_m\}$ (in certain contexts, the elements of $S$ are called \textit{spins}). A configuration of the system is a feasible vector $\v x\in S^n$; when all particles are independent, $S^n$ is the sets of all configurations. 

We have in mind a new type of statistical mechanics, where a configuration of the $n$ particle system is represented by  a flag of vector spaces $V_1 \subset V_2 \subset ... \subset V_m = \Ff_q^n$. 

In the classical case of independent particles, the total energy of a configuration $\v x$ just depends on its type $(k_i)_{1\leq i \leq m}$, where $k_i$ is the number of appearances of the symbol $s_i$ in $\v x$. In fact, the mean (internal) energy is $\sum_{i=1}^m \frac{k_i}{n} E_i$, where $E_i \in \Rr$ is the energy associated to the spin $s_i$. Setting $E_{m+1}=0$, $\tilde E_i = E_i - E_{i+1}$ and $r_i = \sum_{j=1}^i k_j$, one can write $\sum_{i=1}^n \frac{r_i}{m} \tilde E_i$ instead of $\sum_{i=1}^m \frac{k_i}{n} E_i$.

Now we plan to move beyond independence, so it is convenient to see the energy as a ``global'' function that depends on the type of the sequence. We assume now that the energy associated to a flag of vector spaces  $V_1 \subset V_2 \subset ... \subset V_m = \Ff_q^n$ just depends on its type $(k_1,...,k_m)$ and is of the form 
\begin{equation}
\sum_{i=1}^m \frac{k_i}{n} E_i=\sum_{i=1}^n \frac{r_i}{m} \tilde E_i = \sum_{i=1}^m \frac{(\dim V_i)}{n} \tilde E_i
\end{equation}
where $r_i = \sum_{j=1}^i k_j$, as before. 

In general, if $n>1$, the equations
\begin{align}
\sum_{i=1}^m \frac{k_i}{n} E_i &= \langle E \rangle \label{ME_constraints1}\\
\sum_{i=1}^m k_i &= n, \label{ME_constraints2}
\end{align}
where  $\langle E \rangle\in \Rr$ is a prescribed mean energy, do not suffice to determine the type $(k_1,...,k_m)$ and an additional principle must be introduced to select the ``best'' estimate: the \textit{principle of maximum entropy} \cite{jaynes}. This principle---attributed to Boltzmann and popularized by Jaynes---states that, between all the types that satisfy \eqref{ME_constraints1} and \eqref{ME_constraints2}, we should select the one that corresponds to the greatest number of configurations of the system. This means that we must maximize 
\begin{equation}
W(k_1,...,k_m):= \qbinom{n}{k_1, k_2,...,k_m}
\end{equation}
under the constraints \eqref{ME_constraints1} and \eqref{ME_constraints2}. The maximization of $W(k_1,...,k_m)$ is equivalent to the maximization of $2\log_q W(k_1,...,k_m)/n^2$; as $n\to \infty$, the latter quantity approaches $\ent 2(g_1,...,g_m)$, with $g_i := \lim_n k_i/n$. The maximum entropy principle says that the best estimate to  $(g_1,...,g_m)$ corresponds to the solution to the following problem
\begin{align*}
\max &\quad \ent 2(g_1,...,g_m) \\
\text{subject to} & \quad\sum_{i=1}^m g_i E_j =  \langle E \rangle \\
&  \quad\sum_{i=1}^m g_i = 1.
\end{align*}
This is different from usual presentations of the maximum entropy principle in the literature concerning nonextensive statistical mechanics. Usually the constraints are written in terms of escort distributions derived from $(g_1,...,g_m)$; these have proven useful in several domains, e.g. the analysis of multifractals. However, it is not clear for us how to derive them from combinatorial facts.

\section{Dynamical model}\label{sec:dynamical_model}

When $q$-is a prime power, the $q$-binomial coefficients count vector spaces. As explained in the introduction, this motivates a generalization of information theory where messages are vector spaces in correspondence with the usual information theory for memoryless Bernoulli sources. Table \ref{table_correspondence} outline the correspondence.  Sections \ref{sec:q-binomial-distribution} and \ref{sec:grassmannian_proc} justify the last row of this table. Section \ref{sec:q-binomial-distribution} describes the $q$-deformed version of the binomial distribution, associated to the $q$-binomial coefficients. Section \ref{sec:grassmannian_proc} introduces an original stochastic model for the generation of generalized messages: a discrete-time stochastic process that gives at time $n$ a vector subspace of $\Ff_q^n$. We call it \emph{Grassmannian process}. Finally, Section \ref{sec:asymptotics-grassmannian-process} establishes some facts about the asymptotic behavior of this process.

\subsection{The $q$-binomial distribution}\label{sec:q-binomial-distribution}
Let $Z$ be a random variable that takes the value $1$ with probability $\xi \in[0,1]$ and the value $0$ with probability $1-\xi$ (Bernoulli distribution). Its characteristic function is 
\begin{equation}
\Ee(\e^{itZ}) = \xi \e^{it} + (1-\xi).
\end{equation}
Let $W_n$ be a random variable with values in $\{0,...,n\}$, such that $k$ has probability $\Bin(k|n,\xi): = {n \choose k} \xi^k(1-\xi)^{n-k}$, where $\xi \in[0,1]$. The binomial theorem implies that $\Bin(\cdot|n,\xi)$ is a probability mass function, corresponding to the so-called  binomial distribution. The theorem also implies that
\begin{equation}
\begin{split}
(\Ex{\e^{itZ}})^n &{ }={ } (\xi \e^{it}+(1-\xi))^n \\
&{ }={ } \sum_{k=0}^n {n \choose k} \e^{it k} \xi^k(1-\xi)^{n-k} \\
&{ }={ } \Ex{\e^{itW_n}},
\end{split}
\end{equation}
which means that $W_n = Z_1 + ... + Z_n$ (in law), where $Z_1,...,Z_n$ are $n$ i.i.d. variables with the same distribution than $Z$ \cite[Ch.~I, Sec.~11]{doob}. Given a collection $\{Z_i\}_{i\geq 1}$ of i.i.d. random variables such that $Z_i \sim \Ber(\xi)$, the process $\{W_n\}_{n\geq 1}$ defined by $W_1=Z_1$ and $W_n = W_{n-1}+Z_n$ when $n>1$ is an $\Nn$-valued markovian stochastic process.

There is a well known combinatorial interpretation for all this: if you generate binary sequences of length $n$ by tossing $n$ times a coin that gives $1$ with probability $\xi$ and $0$ with probability $1-\xi$, any sequence with exactly $k$ ones has probability $\xi^k(1-\xi)^{n-k}$ and there are ${n\choose k}$ of them. Therefore, if $Y$ is the sum of the outputs of all the coins (the number of ones in the generated sequence), the probability of observing $Y=k$ is ${n\choose k} \xi^k(1-\xi)^{n-k}$.

There is also a $q$-binomial theorem, known as the Gauss binomial formula \cite[Ch.~5]{kac2001quantum}:
\begin{equation}\label{gauss_binomial_theorem}
(x+y)(x+y q) \cdots (x+y q^{n-1}) = \sum_{k=0}^n \qbinom nk q^{{k\choose 2}}y^k x^{n-k}.
\end{equation}
Let us write $(x+y)^n_q$ instead of $(x+y)(x+y q) \cdots (x+y q^{n-1})$: the $q$-analog of $(x+y)^n$. Then  \eqref{gauss_binomial_theorem} implies that 
\begin{equation}\label{qbinomial-law-xy}
\qBin(k|n,x,y) := \qbinom nk \frac{q^{{k\choose 2}}y^k x^{n-k}}{(x+y)^n_q}
\end{equation}
is a probability mass function for $k\in \{0,...,n\}$, with parameters $n\in \Nn$, $x\geq 0$ and $y \geq 0$. Moreover, the factorization
\begin{equation}\label{factorization_q_binomial_law}
\prod_{j=0}^{n-1} \frac{(x+y\e^{it}q^j)}{(x+yq^j)} = \sum_{k=0}^n \qbinom nk \frac{\e^{itk}y^k x^{n-k}q^{{k\choose 2}}}{(x+y)^n_q}
\end{equation}
shows that a variable $Y_n$ with law $\qBin(n,x,y)$ can be written as the sum of $n$ independent variables $X_1,...,X_n$, such that $X_i$ takes the value $0$ or $1$ with probability $x/(x+yq^{i-1})$ and $yq^{i-1}/(x+yq^{i-1})$, respectively. 

If we begin with a collection $\{X_i\}_{i\geq 1}$ of independent variables such that $X_i\sim \Ber\left(\frac{yq^{i-1}}{x+yq^{i-1}}\right)$, then the process $\{Y_n\}_{n\geq 1}$ defined by $Y_n = X_1 + \cdots + X_n$ is an $\Nn$-valued markovian stochastic process. When $q\to 1$, each $X_i$ becomes a Bernoulli variable with parameter $y/(x+y)$ and $Y$ has a $\Bin(n, \frac{y}{x+y})$ distribution. Equation \eqref{factorization_q_binomial_law} also implies that
\begin{equation}
\Ee(Y) = \sum_{j=0}^{n-1} \frac{yq^j}{x+yq^j} = n - \sum_{j=0}^{n-1} \frac{x}{x+yq^j}.
\end{equation}

Provided that $x\neq 0$, one can write the mass function of the $q$-binomial as follows: 
\begin{equation}\label{qbinomial-law}
\qBin(k|n,\theta) := \qbinom nk \frac{q^{{k\choose 2}}\theta^k}{(-\theta;q)_n},
\end{equation}
where $\theta = y/x \geq 0$. We adopt here the classical notation $(-\theta; q)_n$ instead of $(1+\theta)^n_q$.\footnote{The notation can be misleading, because the terms $1$ and $\theta$ do not commute inside $(1+\theta)^n_q$.} Strictly speaking, this is the $q$-binomial distribution found in the literature  \cite{kemp}. The expectation and the variance of this simplified distribution are respectively
\begin{align}
\Ee(Y) &= \sum_{j=0}^{n-1} \frac{\theta q^j}{1+\theta q^j} = n - \sum_{j=0}^{n-1} \frac{1}{1+\theta q^j},\label{mean} \\
\Var(Y) &= \sum_{j=0}^{n-1} \frac{\theta q^j}{(1+\theta q^j)^2}.
\end{align}
The statistical estimation of $\theta$ is addressed in the Appendix \ref{app:maximum_likelihood}.

Set $c_n(\theta):=\sum_{j=0}^{n-1} \frac{1}{1+\theta q^j}$; this sequence is monotonic in $n$ and convergent to certain $c(\theta)$. We do not include $q$ in the notation, since it is fixed from the beginning.

\subsection{A vector-space valued stochastic process associated to the $q$-binomial distribution}\label{sec:grassmannian_proc}
 The vector $(Z_1,...,Z_n)$ is a random binary sequence, but its $q$-deformation $(X_1,...,X_n)$, obtained in the previous section, cannot be identified in an obvious way with a vector space. This motivates the introduction of an associated stochastic process $\{V_i\}_{i\in \Nn}$ such that, for each $n\in \Nn$, $V_n$ is vector subspace of $\Ff_q^n$ and the law of $\{X_i\}_{i\in \Nn^*}$ can be recovered from that of $\{V_i\}_{i\in \Nn}$.

 Let $\Gr(k,n)$ be the set of $k$-dimensional vector subspaces of $\Ff_q^n$ and define the total $n$-th Grassmannian by
\begin{equation}
\Gr(n):= \bigcup_{i=0}^n \Gr(i,n).
\end{equation} 
\begin{definition}[Grassmannian process] Let $\langle 0\rangle=\Ff_q^0 \hookrightarrow \Ff_q^1 \hookrightarrow \Ff_q^2 \hookrightarrow .... \hookrightarrow \Ff_q^n \hookrightarrow...$ be a sequence of linear embeddings; note that it induces embeddings at the level of Grassmannians, that will be implicit in what follows. Define $V_0 := \Ff_q^0$, the trivial vector space; for each $n\geq 0$, let $V_{n+1}$ be a random variable taking values in $\Gr(n+1)$ with law defined by
\begin{align}
\Pr{V_{n+1}  =v | V_n = w, X_{n+1} = 0} &= \delta_w(v) ,\\
\Pr{V_{n+1} =v | V_n = w, X_{n+1} = 1} &= \frac{[v\in \Dil_{n+1}(w)]}{|\Dil_{n+1}(w)|}.
\end{align}
The $(n+1)$-dilations of $w$, $\Dil_{n+1}(w)$, are defined as  $$\set{v \in \Gr(n+1)}{w\subset v, \: v\not\subset \Ff_q^n, \: \dim v - \dim w = 1}.$$ We shall refer to $\{V_n\}_{n\in \Nn}$ as the Grassmannian process associated to the $q$-binomial process.
\end{definition}

\begin{proposition}\label{law_Vn} Let $v$ be a subspace of $\Ff_q^n$ such that $\dim(v) = k$. Then,
\begin{equation}\label{pmf_vector_spaces}
\Pr{V_n = v }=\frac{\theta^kq^{k(k-1)/2}}{(-\theta;q)_n}.
\end{equation}
\end{proposition}
\begin{IEEEproof} See Appendix \ref{proof_law_Vn}
\end{IEEEproof}

\begin{corollary}
\begin{equation}
\Pr{\dim V_n =k} = \qbinom nk \frac{\theta^kq^{k(k-1)/2}}{(-\theta;q)_n}.
\end{equation}
\end{corollary}
\begin{IEEEproof}
This is a consequence of Proposition \ref{law_Vn} and the fact that $\qbinom nk$ counts the number of $k$ dimensional subspaces of $\Ff_q^n$.
\end{IEEEproof}

\begin{proposition}\label{prop:simplified_law_Vn} Let $\{Y_n\}_{n\in \Nn^*}$ denote a $q$-binomial process, $Y_n \sim \Bin_q(n,\theta)$, and $\{V_n\}_{n\in \Nn}$ its associated Grassmannian process. Let $v$ be a subspace of $\Ff_q^n$ of dimension $k=n-d$, for $d \in \llbracket 0, n\rrbracket$.  Then,
\begin{equation}\label{proba_Vn_for_d}
\Pr{V_n = v} = \frac{q^{-\frac 12 (d -(\frac 12 - \log_q\theta))^2 + \frac 12 (\frac 12 - \log_q\theta)^2 - \frac{n^2}{2} \ent 2(d/n) } }{(-\theta^{-1}; q^{-1})_n}.
\end{equation}
\end{proposition}
\begin{IEEEproof}
We shall rewrite the various factors in \eqref{pmf_vector_spaces}. In the first place,
\begin{equation}
(-\theta;q)_n = \prod_{i=0}^{n-1} \theta q^i (1+\frac{1}{\theta q^i}) = \theta^n q^{n(n-1)/2} (-\theta^{-1};q^{-1})_n. 
\end{equation}
Note also that $n^2 \ent 2(d/n) = n^2 - k^2 -d^2$, which implies
\begin{align}
q^{{k\choose 2}} = q^{k^2/2}q^{-k/2}  
&= q^{(n^2-n^2\ent 2(d/n) -d^2)/2 } q^{(d-n)/2}.
\end{align}
Finally, $\theta^k = \theta^{n-d}$. Replace all this in \eqref{pmf_vector_spaces} and simplify to obtain
\begin{equation}
\Pr{V_n = v} = \frac{q^{-\frac{d^2}{2} + d (\frac 12 - \log_q\theta)} q^{-\frac{n^2}{2} \ent 2(d/n)}}{(-\theta^{-1}; q^{-1})_n}.
\end{equation}
Complete the square in the exponent to conclude.
\end{IEEEproof}

\subsection{Asymptotics}\label{sec:asymptotics-grassmannian-process}
Let us define a function $\mu:\Nn\to (0,\infty)$ by
\begin{equation}
\mu(d) := \frac{q^{-\frac 12 (d -(\frac 12 - \log_q\theta))^2 + \frac 12 (\frac 12 - \log_q\theta)^2 } (q^{-(d+1)};q^{-1})_\infty}{(q^{-1};q^{-1})_\infty (-\theta^{-1}; q^{-1})_\infty},
\end{equation}  
and introduce the notation $\mu(\llbracket a, b\rrbracket):=\sum_{d=a}^b \mu(d)$.

 The asymptotic formula in Theorem \ref{prop:asymptotic_behavior}, combined with Proposition \ref{prop:simplified_law_Vn}, implies that
\begin{equation}
\Pr{V_n \in \Gr(n-d,n)}=\qbinom n{n-d} \Pr{V_n=\Ff_q^{n-d}} \to \mu(d),
\end{equation}
for each fixed $d\in \Nn$. 

\begin{proposition}\label{mu_proba}
\begin{equation}
\sum_{d=0}^\infty \mu(d) = 1.
\end{equation}
\end{proposition}
\begin{IEEEproof}
See Appendix \ref{proof_mu_proba}.
\end{IEEEproof}

Therefore, there is a well defined function $ \Delta:[0,1)\to \Nn$ that associates to each $p\in [0,1)$ the smallest $d$ such that $\mu(\llbracket 0, d\rrbracket) \geq p$; explicitly 
\begin{equation}
\Delta(p) = \sum_{k=0}^\infty [p>\mu(\llbracket 0, k\rrbracket)].
\end{equation}
The sum is finite for every $p\in [0,1)$. Note that $ \Delta$ is \emph{left} continuous. This function plays an important role in the proof of Theorem \ref{them:AEP}.

\section{Generalized information theory}\label{sec:generalized_info}

In this section, we prove a fundamental result on measure concentration for the Grassmannian process (Theorem \ref{them:AEP}), that generalizes the asymptotic equipartition property to this setting. It justifies the definition of ``typical subspaces''. Section \ref{sec:coding} applies this result to source coding.

\subsection{Remarks on measure concentration}
The following definition covers the different stochastic models discussed so far.  We use it to clarify the correspondence between Shannon's information theory for sequences and our version for vector subspaces from the probabilistic viewpoint.

\begin{definition}[Refinement of a law] Let $\pi:(A, \mathcal A) \to (B,\mathcal B)$ be a surjection of measurable spaces and $p$ a probability measure on $(B,\mathcal B)$. The law has a refinement with respect to $\pi$ (or $\pi$-refinement) whenever there exists a probability distribution $\tilde p$ on $(A, \mathcal A)$ such that $\pi_* \tilde p = p$, where $\pi_* \tilde p$ denotes the image law (the push-forward of $\tilde p$, its marginalization). 

\end{definition}
In applications, $p$ is the law of a $(B,\mathcal B)$-valued random variable $X$ and $\tilde p$, the law of a $(A, \mathcal A)$-valued random variable $Y$. When $B\subset \Cc$,
\begin{equation}
\Ee_{\tilde p}(\e^{it\pi(Y)})=\Ee_p(\e^{itX}).
\end{equation}

There are four fundamental examples: 
\begin{enumerate}
\item\label{exam_refinement_binomial} The probability measure $\Ber(\xi)^{\times n}$ on $\{0,1\}^n$, that assigns to every sequence with $k$ ones the probability $\xi^k(1-\xi)^{n-k}$, is a refinement of the law $\Bin(n,\xi)$ with respect to the surjection $\pi_1:\{0,1\}^n \to \{0,1,...,n\}, (x_1,...,x_n)\mapsto \sum_i x_i$. 
\item The previous example generalizes to the so-called multinomial distribution. Let $S=\{s_1,...,s_m\}$ be a finite set and $\mu$ any probability law on $S$; set $p_i:=\mu(\{s_i\})$. The law $\mu^{\otimes n}$ assigns to a sequence $x$ in $S^n$ the probability $\prod_{i=1}^m p_i^{a_i(x)}$, where $a_i(x)$ denotes the number of appearances of the symbol $s_i$ in the sequence $x$. Let $T=\set{(k_1,...,k_m)\in \Nn^m}{\sum_{i=1}^m  k_i= n}$; there is a surjection $\pi_2:S^n\to T$ given by $x\mapsto (a_1(x),...,a_m(x))$. Denote by $\nu$ the marginalization of $\mu^{\otimes n}$ under this map, given explicitly by $\nu(\{(k_1,...,k_m)\}) = { n \choose k_1,...,k_n } \prod_{i=1}^m p_i^{k_i}$. Then $\mu^{\otimes n}$ is a $\pi_2$-refinement of $\nu$.
\item The probability measure $\prod_{i=1}^{n-1} \Ber(\frac{\theta q^i}{1+\theta q^{i}})$ on $\{0,1\}^n$ is a refinement of the law $\Bin_q(n,\theta)$ under the application $\pi_1$ introduced above, see  \eqref{factorization_q_binomial_law}.
\item The probability measure on $\Gr(n)$ defined by  \eqref{pmf_vector_spaces}, that we denote $\Grass (n,\theta)$, is also a refinement of $\Bin_q(n,\theta)$ with respect to the surjection $\pi_3:\Gr(n)\to \{0,1,...,n\}, V\mapsto\dim V$.
\end{enumerate}

Let us consider for a moment the binomial case \ref{exam_refinement_binomial}. For $W_n\sim \Bin(n,p)$, Chebyshev's inequality reads $\Pr{ |W_n-pn| >n^{\frac 12+\xi}}\leq p(1-p)/n^{2\xi}$, which goes to $0$ as long as $\xi>0$. In other words, the measure $\Bin(n,p)$ concentrates on the interval $I_{n,\xi}=\llbracket np-n^{\frac 12 + \xi}, np+n^{\frac 12 + \xi}\rrbracket \cap \llbracket 0, n \rrbracket$, in the sense that $\Pr{W_n\in I_n^c} \to 0$ as $n\to \infty$, and therefore the measure $\Ber(\xi)^{\times n}$ concentrates on $\pi_1^{-1}(I_{n,\xi})$, that can be regarded as a set of ``typical sequences''. Moreover, the different type classes $\pi^{-1}(t)$, for $t\in I_{n,\xi}$, have cardinality $\exp\{nH_1(p)+o(n)\}$. An analogous argument shows that the measure $\Bin_q (n,\theta)$ concentrates on the interval $J_{n,\xi}=\llbracket k_n^* - n^\xi, k_n^* + n^\xi \rrbracket \cap \llbracket 0, n \rrbracket$ around the mean $k_n^*$, for any $\xi > 0$, and hence $\Grass (n,\theta)$ concentrates on $\pi_3^{-1}(J_{n,\xi})$. However, there is a difference: while $\Bin(k|n,p)$ goes to $0$ for any value of $k$, and in fact on needs more than $\sqrt{n}$ different types $k$ to accumulate asymptotically a prescribed probability $p_\epsilon := 1-\epsilon$, the values of $\Grass (k|n,\theta) = \Pr{V_n\in \Gr(k.n)}$ tend to the constant value $\mu(d)$, independent of $n$. In the limit, only a finite number of different types $k$ are necessary to accumulate probability $p_\epsilon$, and the corresponding type classes differ in size (even asymptotically). Theorem \ref{them:AEP} bellow reflects this particular situation. 

\subsection{Typical subspaces}

We are ready to prove the main result of this article, which extends Theorems 3 and 4 of Shannon's seminal article \cite{shannon1948} to this setting.

\begin{theorem}\label{them:AEP}
Let $\{Y_n\}_{n\in \Nn^*}$ denote a $q$-binomial process, $Y_n \sim \Bin_q(n,\theta)$; $\{V_n\}_{n\in \Nn}$ its associated Grassmannian process; and  $\delta \in (0,1)$ an arbitrary number. Let $\epsilon>0$ be such that  $p_\epsilon :=1-\epsilon$ is a continuity point of $ \Delta$. Define  $A_n=\bigcup_{k=0}^{a_n} \Gr(n-k,n)$ as the smallest set of the form $\bigcup_{k=0}^m \Gr(n-k,n)$ such that $\Pr{V_n \in A_n^c} \leq \epsilon$. Then, there exists $n_0\in \Nn$ such that, for every $n\geq n_0$,
\begin{enumerate}
\item\label{value_Delta} $A_n = \bigcup_{k=0}^{ \Delta(p_\epsilon)} \Gr(n-k,n)$;
\item\label{approx_proba} for any $v\in A_n$ such that $\dim v = k$,
\begin{equation}\label{approx_proba_eq}
\left| \frac{\log_q (\Pr{V_n = v}^{-1})}{n} - \frac{n}{2}\ent 2(k/n) \right| \leq \delta.
\end{equation} 
\end{enumerate}
The size of $A_n$ is optimal, up to the first order in the exponential: let $s(n,\epsilon)$ denote $\min\set{|B_n| }{B_n\subset \Gr(n) \text{ and } \Pr{V_n \in B_n} \geq 1-\epsilon}$;  then
\begin{equation}\label{size_optimality}
\begin{split}
\lim_n \frac{1}{n} \log_q|A_n| & = \lim_n \frac{1}{n} \log_q s(n,\epsilon) \\
& = \lim_n \frac{n}{2} \ent 2(\Delta(p_\epsilon)/n) \\
& = \Delta(p_\epsilon). 
\end{split}
\end{equation}
\end{theorem}
The set $A_n$ correspond to the ``typical subspaces'', in analogy with typical sequences. 

\begin{IEEEproof}
To shorten the notation, let us write $\Prr n A$ instead of $\Pr{V_n\in A}$, and  $G_{k}^n$ instead of $\Gr(k,n)$.

Given any $\eta>0$, there exists $n(\eta)\in \Nn$ such that, for every $n\geq n(\eta)$ and every $d\in \llbracket 0, \Delta(p_\epsilon)\rrbracket$,
\begin{equation}
|\Prr n{ G_{n-d}^n} - \mu(d)| < \frac{\eta}{ \Delta(p_\epsilon) +1},
\end{equation}
 because $\Prr n {G_{n-d}^n} \to \mu(d)$ for each $d$. 

Since $p_\epsilon$ is a continuity point of $\Delta$, a piece-wise constant function, there exists $\xi>0$ such that $$\Delta(1-\epsilon - \xi) = \Delta(1-\epsilon) = \Delta(1-\epsilon + \xi).$$  Remark now that, for every $n\geq n(\xi)$,
\begin{equation}\label{bound_proba_A_n_Delta_inf}
\sum_{d=0}^{\Delta(p_\epsilon)} \Prr n {G_{n-d}^n} > \sum_{d=0}^{\Delta(p_\epsilon)} \mu(d) -\xi \geq 1-\epsilon,
\end{equation}
because $\mu(\llbracket 0, \Delta(p_\epsilon)\rrbracket) = \sum_{d=0}^{\Delta(p_\epsilon)} \mu(d) \geq 1-\epsilon + \xi$. This is a direct consequence of $\Delta(p_\epsilon) = \Delta(1-\epsilon + \xi)$.

 Analogously, for each $n\geq n(\xi)$,
\begin{align}
\sum_{d=0}^{\Delta(p_\epsilon)-1} \Prr n { G_{n-d}^n} & < \sum_{d=0}^{\Delta(p_\epsilon)-1} \mu(d) + \frac{\Delta(p_\epsilon)}{\Delta(p_\epsilon) + 1} \xi \nonumber\\ 
& < 1-\epsilon- \frac{\xi}{\Delta(p_\epsilon) + 1}  \nonumber\\ 
& < 1-\epsilon,\label{bound_proba_A_n_Delta_sup}
\end{align}
because $\mu(\llbracket 0, \Delta(p_\epsilon)-1\rrbracket) < 1-\epsilon - \xi$: if this is not the case, $\Delta(1-\epsilon - \xi)\leq \Delta(\epsilon)-1$. The inequalities \eqref{bound_proba_A_n_Delta_inf} and \eqref{bound_proba_A_n_Delta_sup} imply the part \ref{value_Delta} of the theorem whenever $n \geq n(\xi)$.

 We suppose now that $n>n(\xi)$. Let $v$ be an element of $A_n$ of dimension $k$, and set $d = n-k$. The formula in Proposition \ref{prop:simplified_law_Vn} can be stated as
\begin{equation}
- \frac{\log_q \Pr{V_n = v}}{n} = \frac{g(d,n)}{n}   + \frac{n}{2} \ent 2(d/n),
\end{equation}
where we have set $g(d,n)= \frac 12 (d -(\frac 12 - \log_q\theta))^2 -  \frac 12 (\frac 12 - \log_q\theta)^2 +  \log_q (-\theta^{-1}; q^{-1})_n$.  Since $d$ belongs to the interval $\llbracket 0, \Delta(p_\epsilon)\rrbracket$, independent on $n$, and $(-\theta^{-1}; q^{-1})_n \to (-\theta^{-1}; q^{-1})_\infty$, there exists $n_0 \geq n(\xi)$ such that, for every $n\geq n_0$ and every $d\in \llbracket 0, \Delta(p_\epsilon)\rrbracket$, $g(d,n)/n <\delta,$ which proves part \ref{approx_proba} of the theorem. 

For $n$ big enough, $\Delta(p_\epsilon)$ belongs to the interval $[n/2,n]$. The inequalities in Lemma \ref{lemma:unimodality_binomial} imply that
\begin{equation}\label{inequalities_size_An}
\begin{split}
\qbinom n {n-\Delta(p_\epsilon)} &\leq |A_n| \\
&\leq \sum_{k=0}^{\Delta(p_\epsilon)} \qbinom n {n-k} \\
&\leq (\Delta(p_\epsilon)+1) \qbinom n {n-\Delta(p_\epsilon)}.
\end{split}
\end{equation}
Therefore,
\begin{equation}\label{asymptotic_size_An}
\lim_n \frac 1n \log_q |A_n|  = \lim_n \frac 1n \log_q \qbinom n {n-\Delta(p_\epsilon)} = \Delta(p_\epsilon),
\end{equation}
where the second equality comes from \eqref{computation_symptotics_binomial}.

For any $\epsilon$, we show now how to build iteratively a set $B_n$ of minimal cardinality such that $\Prr n {B_n^c} \leq \epsilon$: start with $B_n = \emptyset$ and then add vector subspaces of $\Ff_q^n$ one-by-one, picking at each time any of the vector subspaces of \emph{highest dimension} in $B_n^c$, until you attain  $\Prr n {B_n^c} \leq \epsilon$. Let $n-b_n$ be the dimension of the last space included in $B_n$. It is easy to prove that $b_n < 2\sqrt{n}$, as a consequence of Chebyshev's inequality (the interval $[n-2\sqrt{n},n]$ accumulates probability $p_\epsilon$ when $n$ is big enough).  This construction gives in fact the smallest possible set, because the function $f_n:[0,n] \to \Rr, \: x \mapsto \theta^x q^{x(x-1)/2}/(-\theta,q)_n$ is strictly convex and  attains its minimum at $x_0=\frac 12 - \log_q\theta$; therefore, all the subspaces are included in $B_n$ in decreasing order of probability, and the probability of the last space included is bounded bellow by $\theta^{n-2\sqrt n} q^{({n-2\sqrt n})({n-2\sqrt n}-1)/2}/(-\theta,q)_n$, which is  much bigger that $(-\theta,q)_n^{-1}$, the maximum of $f_n$ on $[0,x_0]$, when $n$ is big enough.

Two versions of $B_n$ only differ in the particular subspaces of dimension $n-b_n$ they include, but they coincide on $\bigcup_{k=0}^{b_n-1} G_{n-k}^n$. In what follows, $B_n$ denotes any of the possible sets. Remark also that $B_n \subset A_n$; even more, $a_n=b_n$ (a strict inequality between the two contradicts the minimality of either $B_n$ or $a_n$). It is also true in general that 
\begin{align}
 p_\epsilon &\leq \Prr n { B_n} \nonumber\\
 & = \sum_{k=0}^{a_n} \Prr n { B_n\cap G_{n-k}^n} \nonumber\\
 & = \Prr n {B_n\cap G_{n-a_n}^n} + \sum_{k=0}^{a_n-1} \Prr n {B_n\cap G_{n-k}^n}.\label{inequality_probas_B_n}
\end{align}

We restrict ourselves again to the case in which $p_\epsilon$ is continuity point of $\Delta$, in such a way that $\Delta(p_\epsilon)= a_n=b_n$. Under these hypotheses, we are able to lower-bound uniformly the term $\Prr n {B_n\cap G_{n-\Delta(p_\epsilon)}^n}$ by using \eqref{inequality_probas_B_n}, and deduce from this that $|B_n|$ grows like $|A_n|$, that in turn grows like $|G_{n-\Delta(p_\epsilon)}^n|$, as shown in \eqref{asymptotic_size_An}. In fact, we have that
\begin{align}
\sum_{k=0}^{\Delta(p_\epsilon)-1} \Prr n {  B_n\cap G_{n-k}^n} &\leq \sum_{k=0}^{\Delta(p_\epsilon)-1} \Prr n {G_{n-k}^n} \nonumber \\ 
& { }<{ } 1 - \epsilon - \frac{\xi}{\Delta(p_\epsilon)+1}, \label{bound_B_n_inter_grassmanian_2} 
\end{align}
where we have used again the bound in \eqref{bound_proba_A_n_Delta_sup}. Inequalities \eqref{inequality_probas_B_n} and \eqref{bound_B_n_inter_grassmanian_2} imply that
\begin{equation}
\frac{\xi}{\Delta(p_\epsilon)+1} <  \Prr n {B_n\cap G_{n-\Delta(p_\epsilon)}^n}.
\end{equation}
When $n>n_0$, the part \eqref{approx_proba} entails that $\Prr n x \leq q^{-n^2 \ent 2(\Delta/n)/2 + n\delta}$ for every $x\in G_{n-\Delta(p_\epsilon)}^n$, or equivalently  $\Prr n x q^{n^2 \ent 2(\Delta/n)/2 - n\delta} \leq 1$. Then,
\begin{align}
|B_n| & \geq |B_n \cap  G_{n-\Delta(p_\epsilon)}^n| \nonumber\\ 
&  \geq \sum_{x\in B_n \cap G_{n-\Delta(p_\epsilon)}^n} \Prr n x q^{n^2 \ent 2(\Delta(p_\epsilon)/n)/2 - n\delta} \nonumber\\
& \geq q^{n^2 \ent 2(\Delta(p_\epsilon)/n)/2 - n\delta} \Prr n {B_n \cap G_{n-\Delta(p_\epsilon)}^n} \nonumber\\
& > q^{n^2 \ent 2(\Delta(p_\epsilon)/n)/2 - n\delta} \frac{\xi}{\Delta(p_\epsilon)+1}.\label{lower_bound_Bn_with_probas}
\end{align}
We deduce that 
\begin{equation}\label{lim_inf_Bn}
\liminf_n \frac 1n \log_q |B_n| \geq \lim_n \frac{n}{2} \ent 2(\Delta(p_\epsilon)/n) - \delta.
\end{equation}
On the other hand, since $B_n \subset A_n$, it is clear that 
\begin{equation}\label{lim_sup_Bn}
\begin{split}
\limsup \frac 1n \log_q |B_n| &\leq \lim_n \frac 1n \log_q |A_n| \\
&{ }={ } \lim_n \frac{n}{2} \ent 2(\Delta(p_\epsilon)/n).
\end{split}
\end{equation}
Since $\delta>0$ is arbitrarily small,  \eqref{lim_inf_Bn} and \eqref{lim_sup_Bn} imply that $\lim_n \frac 1n \log_q |B_n|$  exists and equals $\Delta(p_\epsilon)$. The theorem is proved.
\end{IEEEproof}

\begin{remark}
The definition of  $A_n$ still makes sense when $p_\epsilon$ is a discontinuity point of $ \Delta$. In this case, there exists $\xi>0$ such that $\Delta(p_\epsilon)+1 = \Delta(p_\epsilon + \xi)$ and $\Delta(p_\epsilon)= \Delta(p_\epsilon-\xi)$ . Inequality \eqref{bound_proba_A_n_Delta_inf} can  be easily adapted to show that $\sum_{k=0}^{\Delta(p_\epsilon)+1} \Gr(n-k,n) \geq 1-\epsilon$, which implies that $a_n\leq \Delta(p_\epsilon)+1$; by  \eqref{bound_proba_A_n_Delta_sup}, $a_n\geq \Delta(p_\epsilon)$. Of course, part \ref{approx_proba} in the Theorem still makes sense. We also have that $B_n\subset A_n$ and $a_n = b_n$. The problems appear in the comparison of $|B_n|$ and $|A_n|$; it is possible that $\Prr n {B_n\cap \Gr(n-\Delta(p_\epsilon),n)}$ goes to zero very fast when $n\to \infty$, and  \eqref{lim_inf_Bn} is not valid any more. However, we can still adapt the bounds in \eqref{lower_bound_Bn_with_probas} to prove
\begin{equation*}
\begin{split}
\liminf_n \frac 1n \log_q |A_n|  & { }\geq{ } \liminf_n \frac 1n \log_q |B_n| \\
&{ }\geq{ } \lim_n \frac 1n \log_q \qbinom n {n-(\Delta(p_\epsilon)-1)} \\
&{ }={ } \Delta(p_\epsilon)-1,
\end{split}
\end{equation*}
because $b_n = a_n \geq \Delta(p_\epsilon)$ and therefore $ \Gr(n-(\Delta(p_\epsilon)-1),n)\subset B_n$. Analogously, $B_n\subset A_n$ and $a_n \leq  \Delta(p_\epsilon)+1$ lead to
\begin{equation*}
\begin{split}
\limsup_n \frac 1n \log_q |B_n| & { }\leq{ } \limsup_n  \frac 1n \log_q |A_n| \\
&{ }\leq{ } \lim_n  \frac 1n \log_q \qbinom n {n-(\Delta(p_\epsilon)+1)} \\
&{ }={ } \Delta(p_\epsilon)+1,
\end{split}
\end{equation*}
where we have used again \eqref{inequalities_size_An}.
\end{remark}

\begin{remark}
In the classical case of sequences, all the typical sequences tend to be equiprobable, in the sense of  \eqref{eq:class_AEP_equiprobability}. This is not valid for the process $V_n$: a typical space $v\in A_n$ of dimension $n-d$ satisfy asymptotically the bounds $q^{-n(\frac n2 \ent 2(d/n)+\delta)} \leq \Pr{V_n = v}\leq q^{-n(\frac n2 \ent 2(d/n)-\delta)}$, for any $\delta>0$, and  $\frac n2 \ent 2(d/n) = d + O(1/n)$. 
\end{remark}

\subsection{Coding}\label{sec:coding}

Inspired by \cite{csiszar1981information}, we define a generalized $n$-to-$k$ $q$-ary block code as a pair of mappings $f:\Gr(n)\to \{1,...,q\}^k$ and $\phi:\{1,...,q\}^k\to \Gr(n)$. For a given stochastic process $W_n$, such that $W_n$ takes values in $\Gr(n)$, we define the probability of error of this code as $e(f,\phi)=\Pr{\phi(f(W_n))\neq W_n}$. Small $k$ and small probability of error are good properties for codes, but there is a trade-off between the two.  Let $k(n,\epsilon)$ be the smallest $k$ such that there exists a generalized $n$-to-$k$ $q$-ary block code $(f,\phi)$ that satisfies $e(f,\phi)\leq \epsilon$. 
\begin{proposition}\label{minimum_block_code}
For the Grassmanian process $V_n$ introduced above and for all $\epsilon>0$ such that $p_\epsilon=1-\epsilon$ is a continuity point of $\Delta$, one has
\begin{equation}
\lim_n \frac{k(n,\epsilon)}{n}= \Delta(p_\epsilon).
\end{equation}
\end{proposition}
\begin{IEEEproof}
The existence of an $n$-to-$k$ $q$-ary block code $(f,\phi)$ such that $e(f,\phi)\leq \epsilon$ is equivalent to the existence of a set $B_n\subset \Gr(n)$ such that $\Pr{V_n\in B_n}\geq 1-\epsilon$ and $|B_n|\leq q^k$ (let $B_n$ be the set of sequences that are reproduced correctly...). As in the main theorem, let $s(n,\epsilon)$ denote the minimum cardinality of such a set. The statement in Proposition \ref{minimum_block_code} is therefore equivalent to $\lim_n \frac 1n \log_q s(n,\epsilon) = \Delta(p_\epsilon)$, which is already proved.
\end{IEEEproof}

In simpler terms, it is always possible to code all the typical subspaces $A_n=\bigcup_{k=0}^{ \Delta(p_\epsilon)} \Gr(n-k,n)$ with different code-words if one disposes of $q^{n(\Delta(p_\epsilon) +\xi)}$ such words, for  $\xi$ positive and arbitrarily small, as long as $n$ is big enough. In contrast, it is asymptotically impossible if one disposes of $q^{n(\Delta(p_\epsilon) -\xi')}$ different code-words, for any $\xi'>0$.

\section{Further remarks}

A recent paper \cite{jensen2016statistical} proposes the study of ``exploding'' phase spaces: statistical systems such that the cardinality of the space of configurations grows faster than $k^n$, the combination of $n$ components that can occupy $k$ states. The total grassmannians $\Gr(n)=\Gr(n, \Ff_q)$ are an example, since their cardinality grows like $q^{\frac{n^2}{4}+o(n^2)}$. This can be deduced from the unimodality of the $q$-binomial coefficients (Lemma \ref{lemma:unimodality_binomial}) and our asymptotic formulae, because
\begin{equation}
\qbinom n{\floor{n/2}} \leq |\Gr(n)| \leq (n+1)\qbinom n{\floor{n/2}} 
\end{equation} 
and therefore
\begin{equation}
\begin{split}
\lim_n \frac{2}{n^2} \log_q |\Gr(n)| &{ }={ } \lim_n \frac{2}{n^2} \log_q \qbinom n{\floor{n/2}}  \\
&{ }={ } \ent 2\left(\frac 12,\frac 12\right) = \frac 12.
\end{split}
\end{equation}
In fact, the values of $\lim_{n\to\infty} |\Gr(2n+1)|q^{-(2n+1)^2/4}$ and $\lim_{n\to\infty} |\Gr(2n)|q^{-(2n)^2/4}$  depend only on $q$ and can be determined explicitly in terms of the Euler's generating function for the partition numbers and the Jacobi theta functions $\vartheta_2$ and $\vartheta_3$, see \cite[Cor. 3.7]{kousidis}

A link between Tsallis entropy and the size of the \emph{effective} phase space (the configurations whose probability is non-zero) is already suggested by Tsallis in \cite[Sec. 3.3.4]{tsallis-book}. There, $H_{(\rho-1)/\rho}$ appears naturally as a extensive quantity when the effective phase space grows like $N^\rho$, for $\rho>0$. 

Finally, we conjecture the existence of other combinatorial  quantities ${{n}\choose {k_1,...,k_s}}_{\mathrm{gen}}$ that satisfy the multiplicative relations \eqref{multiplicative_identity_multinomials}, but such that 
\begin{equation}
{{n}\choose {p_1n,...,p_sn}}_{\mathrm{gen}}\sim \exp(f(p_1,...,p_s)n^\beta + o(n^\beta)).
\end{equation} 
If this is the case, the function $f(p_1,...,p_s)$ would satisfy the functional equation \eqref{cocycle_eqns} for $\alpha=\beta$, and therefore be equal to $KH_\beta$, for an appropriate constant $K$. 

\appendices
\section{Parameter estimation by the maximum likelihood method}\label{app:maximum_likelihood}
Let us suppose we make $n$ independent trials of a variable $Y$ with distribution $\qBin(n,\theta)$, obtaining results $y_1,...,y_m$. The probability of this outcome is 
\begin{equation}
P(y_1,...,y_m|\theta) = \prod_{i=1}^m \qbinom n{y_i} \frac{\theta^{y_i} q^{y_i(y_i-1)/2}}{(-\theta;q)_n}.
\end{equation}
This implies that
\begin{equation}
\difp{\log P}{\theta} = \frac 1\theta \left(\sum_{i=1}^n y_i - m \sum_{j=0}^{n-1} \frac{\theta q^j}{(1+\theta q^j)}\right).
\end{equation}
By the maximum likelihood method, the best estimate for $\theta$, say $\hat \theta$, should maximize $P$ and therefore satisfy $\left.\difp{\log P}{\theta}\right|_{\theta = \hat \theta}=0$; in turn, this equation implies that the empirical mean 
\begin{equation}
 \bar y := \sum_{i=1}^m y_i
\end{equation}
should coincide with the theoretical mean
\begin{equation}
m_{q,n}(\theta) := \sum_{j=0}^{n-1} \frac{\theta q^j}{1+\theta q^j}.
\end{equation}

\begin{proposition}
The map $\theta \mapsto m_{q,n}(\theta) $ establishes a bijection between $[0,\infty)$ and $[0,n)$.
\end{proposition}
If this correspondence is extended by $m_{q,n}(\infty)=n$ |which corresponds to the case $x=0$| the value of  $\hat \theta$ is uniquely determined by the equation $m_{q,n}(\hat \theta) = \bar y$.
\begin{IEEEproof}
Since 
\begin{equation}
\diff{}{\theta}\left(\frac{\theta q^j}{1+\theta q^j}\right) = \frac{q^j}{(1+\theta q^j)^2} > 0,
\end{equation}
$m_{q,n}(\theta)$ is strictly increasing. Moreover, $m_{q,n}(0)= 0$ and $\lim_{\theta \to \infty} m_{q,n}(\theta)=n$.
\end{IEEEproof}

\section{Proof of Proposition \ref{prop:asymptotic_behavior}}\label{proof_prop_asymptotic_q_multinomial}
First, we substitute \eqref{qGamma} in \eqref{q-coeff-qgamma} (the powers of $(q-1)$ cancel):
\begin{multline}\label{first_exp_binom}
\qbinom{n}{k_1,...,k_s} = \\
(q^{-1};q^{-1})_\infty^{1-s} q^{n^2 \ent 2(\frac {k_1}{n},...,\frac{k_s}{n})/2} \frac{\prod_{i=1}^s (q^{-(k_i+1)};q^{-1})_\infty}{ (q^{-(n+1)};q^{-1})_\infty}. 
\end{multline}
Theorem \ref{prop:asymptotic_behavior} is a direct consequence of this equality and the following fact:  for any sequence $\{t_n\}_n$ of positive numbers, 
\begin{equation}\label{eq:limit_pochhammer_1}
\lim_{n\to \infty} (q^{-(t_n +1)};q^{-1})_\infty = 1
\end{equation}
 if $ t_n \to \infty$, and 
 \begin{equation}\label{eq:limit_pochhammer_2}
 \lim_{n\to \infty} (q^{-(t_n +1)};q^{-1})_\infty=(q^{-(t + 1)};q^{-1})_\infty
 \end{equation}
  if $ t_n \to t \in [0,\infty)$.
 
To establish \eqref{eq:limit_pochhammer_1} and \eqref{eq:limit_pochhammer_2},  remark first that $$(q^{-(t_n +1)};q^{-1})_\infty=\sum_{j=0}^\infty q^{-j(t_n +1)}/(q^{-1};q^{-1})_j$$ can be written as $\int_{\Nn} f_n(x) \nu(dx)$, where $\nu$ denotes the counting measure and  $f_n:\Nn\to [0,\infty)$ is given by
\begin{equation}
f_n(x) = \frac{q^{-x(t_n +1)}}{(q^{-1};q^{-1})_x}
\end{equation}
Moreover, $|f_n(x)| \leq g(x):=q^{-x}/(q^{-1};q^{-1})_x$, because $t_n \geq 0$, and $g(x)$ is integrable, $\int_\Nn g(x) \nu(dx) \leq (q^{-1},q^{-1})_\infty^{-1} \frac{1}{1-q^{-1}}$. Therefore, in virtue of Lebesgue's dominated convergence theorem,
\begin{align*}
\lim_{n\to \infty} \sum_{j=0}^\infty \frac{q^{-j(t_n +1)}}{(q^{-1};q^{-1})_j} &= \lim_n \int_{\Nn} f_n(x) \nu(dx) \\
& =\int_{\Nn}  \lim_n  f_n(x) \nu(dx) 
\end{align*}
The point-wise limit $\lim_n  f_n(x)$ is  $[x=0]$ when $t_n\to \infty$ and $ \frac{q^{-x(t +1)}}{(q^{-1};q^{-1})_x}$ when $t_n\to t$. 

\section{Proof of Proposition \ref{law_Vn} }\label{proof_law_Vn} 

To shorten notation, we write in this section $\Prr{X}x$ instead of $\Pr{X=x}$, and $\Prr{X|Y}{x|y}$ instead of $\Pr{X=x|Y=y}$.

Our proof is by recurrence. The case $n=1$ is straightforward; for instance,
\begin{align*}
\Prr{V_1}{\langle 0\rangle} &= \Prr{V_1|V_0}{\langle 0 \rangle| \langle 0 \rangle} \\
&= \Prr{V_1|V_0,X_1}{\langle 0 \rangle| \langle 0 \rangle, 0}\Prr{X_1}{0}, \\
&= \Prr{X_1}{0}
\end{align*}
because $\langle 0 \rangle$ it is not a dilation of itself.

 Suppose the formula is valid up to $n\geq 1$. Let $v'$ be a subspace of $\Ff_q^{n+1}$ of dimension $k$.  When $v'$ is contained in $\Ff_q^n$,
 \begin{align*}
 \Prr{V_{n+1}}{ v' } &= \Prr{V_{n+1}|V_n,X_{n+1} }{v' |  v', 0} \Prr{X_{n+1}}{0}\Prr{V_n}{ v'} \\
 & = 1 \cdot \frac{1}{1+\theta q^n} \frac{\theta^kq^{k(k-1)/2}}{(-\theta;q)_n} = \frac{\theta^k q^{k(k-1)/2}}{(-\theta;q)_{n+1}}.
 \end{align*}
 If $v'\not \subset \Ff_q^n$,
 \begin{IEEEeqnarray*}{l}
 \Prr{V_{n+1}}{ v' } \\
 =  \sum_{w\in \Gr(n)} \Prr{V_{n+1}|V_n,X_{n+1}}{v'|w, 1} \Prr{Y_n}{ w} \Prr{X_{n+1}}{ 1} \\
 =  \sum_{\substack{w\in \Gr(k-1,n)\\ w\subsetneqq V}} \frac{1}{|\Dil_{n+1}(w)|} \left( \frac{\theta^{k-1} q^{{k-1 \choose 2}}}{(-\theta;q)_n}\right) \frac{\theta q^n}{(1+\theta q^n)}\\
 = \frac{\theta^k q^{{k-1\choose 2}} q^n}{|\Dil_{n+1}(v\cap \Ff_q^n)| (-\theta;q)_{n+1}}.
  \end{IEEEeqnarray*}
 The formula $F(U)+F(V)=F(U+V)+F(U\cap V)$ entails that $v\cap \Ff_q^n$ has dimension $k-1$. Any $w \in \Gr(k-1,n)$ such that  $w\subset v$ must be contained in $v\cap \Ff_q^n$ and have the same dimension, implying that $w = v\cap \Ff_q^n$; this explain the last equality above. 
 
 Finally, let $w$ be a $k-1$ dimensional subspace in $\Ff_q^n$; to dilate it into a $v\in \Gr(k,n+1)\sm\Gr(k,n)$, one must pick a vector $x$ outside $\Ff_q^n$: there are $q^{n+1}-q^n$ of those. However, since $w+\langle x\rangle$ has $q^k$ points and $w$ just $q^{k-1}$, there are $q^k-q^{k-1}$ choices of $x$ that give the same dilation $v$. Therefore, the number of different dilations is 
 \begin{equation}
 \frac{q^{n+1}-q^n}{q^k-q^{k-1}} = q^{n-(k-1)}.
\end{equation}  
In particular, the quantity $|\Dil_{n+1}(v\cap \Ff_q^n)|$ equals $q^{n-(k-1)}$.

\section{Proof of Proposition \ref{mu_proba}}\label{proof_mu_proba}
We prove first a lemma that will be useful in the proof of Proposition \ref{mu_proba}.

\begin{lemma}
For every $n\in \Nn$ and every $d\in [0,n]$,
\begin{equation}\label{upper_bound_quotient}
\frac{(q^{-(n-d+1)};q^{-1})_\infty}{q^{-(n+1)};q^{-1})_\infty} \leq 1.
\end{equation}
Moreover, for every $n\in \Nn$ and every $d\in\llbracket 0,2\sqrt{n}\rrbracket$,
\begin{equation}\label{lower_bound_quotient}
1-c(q)q^{-(\sqrt n +1)^2} \leq \frac{(q^{-(n-d+1)};q^{-1})_\infty}{q^{-(n+1)};q^{-1})_\infty},
\end{equation}
where $c(q)=2(q^{-1};q^{-1})_\infty$.
\end{lemma}
\begin{IEEEproof}
In this proof we use repeatedly the $q$-binomial theorem \eqref{eq:q-binomial-theorem}.
For any $k\in\Nn$, $q^{-k(n+1)}\leq q^{-k(n-d+1)}$, which in turn implies \eqref{upper_bound_quotient}:
\begin{align*}
\frac{1}{(q^{-(n+1)};q^{-1})_\infty} &{ }={ } \sum_{k=0}^\infty  \frac{q^{-k(n+1)}}{(q^{-1};q^{-1})_k}\\
 & { }\leq{ } \sum_{k=0}^\infty  \frac{q^{-k(n-d+1)}}{(q^{-1};q^{-1})_k} \\
 & { }={ } \frac{1}{(q^{-(n-d+1)};q^{-1})_\infty}.
\end{align*}
To prove \eqref{lower_bound_quotient}, first remark that
\begin{align*}
 \frac{1}{(q^{-(n-d+1)};q^{-1})_\infty} & - \frac{1}{(q^{-(n+1)};q^{-1})_\infty} \\
 & { }={ } \sum_{k=1}^\infty \frac{q^{-k(n+1)}(q^{kd}-1)}{(q^{-1};q^{-1})_k} \\
& { }\leq{ }  (q^{-1};q^{-1})_\infty^{-1} \sum_{k=1}^\infty q^{-k(n+1)}q^{kd} \\
& { }\leq{ }  (q^{-1};q^{-1})_\infty^{-1} \sum_{k=1}^\infty q^{-k(\sqrt n+1)^2}. 
\end{align*}
Remark that we omit the term corresponding to $k=0$, since it vanishes. The first of these inequalities is implied by the trivial bound $x-1\leq x$ and the fact that $\{(q^{-1};q^{-1})_k\}_k$ decreases with $k$; the second, from  $ d \leq 2\sqrt{n}$. The geometric series $\sum_{k=1}^\infty q^{-k(\sqrt n+1)^2}$ equals $q^{-(\sqrt n+1)^2} (1-q^{-(\sqrt n+1)^2})^{-1}$, that is upper-bounded by $2q^{-(\sqrt n+1)^2}$, because $q\geq 2$. Hence, we have
 \begin{align*}
 \frac{1}{(q^{-(n-d+1)};q^{-1})_\infty} &- \frac{1}{(q^{-(n+1)};q^{-1})_\infty} \\
 & { }\leq{ } 2(q^{-1};q^{-1})_\infty^{-1} q^{-(\sqrt n+1)^2} \\
 &{ }={ }c(q)q^{-(\sqrt n+1)^2} .
 \end{align*}
 Finally, note that $\frac{1}{(q^{-(n-d+1)};q^{-1})_\infty} = 1 +$ (positive term)$\geq 1$, therefore it is also true that 
 \begin{equation}
 \begin{split}
 \frac{1}{(q^{-(n-d+1)};q^{-1})_\infty} & - \frac{1}{(q^{-(n+1)};q^{-1})_\infty} \\
 & { }\leq{ } \frac{c(q)q^{-(\sqrt n+1)^2}}{(q^{-(n-d+1)};q^{-1})_\infty}.
 \end{split}
 \end{equation}
\end{IEEEproof}

\begin{IEEEproof}[Proof of Proposition \ref{mu_proba}]
To simplify notation, set 
\begin{equation}
A(d):=-\frac 12 (d -(\frac 12 - \log_q\theta))^2 + \frac 12 (\frac 12 - \log_q\theta)^2.
\end{equation}
and $B_n=(-\theta^{-1};q^{-1})_n^{-1}$.
 Recall from  \eqref{first_exp_binom} that 
\begin{equation}
\qbinom n{n-d} = \frac{q^{n^2 \ent 2(d/n)/2}(q^{-(d+1)};q^{-1})_\infty (q^{-(n-d+1)};q^{-1})_\infty}{(q^{-1};q^{-1})_\infty (q^{-(n+1)};q^{-1})_\infty }.
\end{equation}
This and \eqref{proba_Vn_for_d} give
\begin{align}
  1&{ }={ }  \sum_{d=0}^n \Pr{V_n\in \Gr(n-d,n)}\nonumber\\
& { }={ } B_n\sum_{d=0}^n \frac{q^{A(d)}(q^{-(d+1)};q^{-1})_\infty}{(q^{-1};q^{-1})_\infty} \frac{(q^{-(n-d+1)};q^{-1})_\infty}{(q^{-(n+1)};q^{-1})_\infty}\nonumber\\
& { }\leq{ } B_n \sum_{d=0}^n \frac{q^{A(d)}(q^{-(d+1)};q^{-1})_\infty}{(q^{-1};q^{-1})_\infty}.\label{lower_bound_sum_mu}
\end{align}
At the end we have used the inequality \eqref{upper_bound_quotient}. In turn,  \eqref{lower_bound_sum_mu} implies that
\begin{equation}
(-\theta^{-1};q^{-1})_\infty \leq  \sum_{d=0}^\infty \frac{q^{A(d)}(q^{-(d+1)};q^{-1})_\infty}{(q^{-1};q^{-1})_\infty}
\end{equation}
We shall see that in fact this is an equality, as the proposition claims. Using this time \eqref{lower_bound_quotient}, we obtain
\begin{align*}
1 & \geq  \sum_{d=0}^{\floor{2 \sqrt n}} \Pr{V_n\in \Gr(n-d,n)}  \\
& \geq B_n \sum_{d=0}^{\floor{2 \sqrt n}} \frac{q^{A(d)}(q^{-(d+1)};q^{-1})_\infty}{(q^{-1};q^{-1})_\infty} (1-c(q)q^{-(\sqrt n +1)^2}).
\end{align*}
which is equivalent to
\begin{equation}
\sum_{d=0}^{\floor{2 \sqrt n}} \frac{q^{A(d)}(q^{-(d+1)};q^{-1})_\infty}{(q^{-1};q^{-1})_\infty} \leq  \frac{(-\theta^{-1};q^{-1})_n}{1-c(q)q^{-(\sqrt n +1)^2}}.
\end{equation}
In the limit, 
\begin{equation}
\sum_{d=0}^{\infty} \frac{q^{A(d)}(q^{-(d+1)};q^{-1})_\infty}{(q^{-1};q^{-1})_\infty}  \leq (-\theta^{-1};q^{-1})_\infty.
\end{equation}
and this finishes the proof.
\end{IEEEproof}

\section*{Acknowledgements}
I am very grateful to Matilde Marcolli, who pointed out the combinatorial meaning of the $q$-multinomial coefficients during a conversation we had at CIRM. I also want to thank Daniel Bennequin for his constant encouragement and multiple suggestions.

\bibliographystyle{IEEEtran}
\bibliography{IEEEabrv,bibtex}

\begin{IEEEbiographynophoto}{Juan Pablo Vigneaux Ariztía}
 received the B.Eng.Sc. and the Engineer's degree in industrial engineering 
 from the Pontifical Catholic University of Chile (PUC), Santiago, Chile, in 2014 
 and the master's degree in fundamental mathematics from Pierre and Marie Curie University (Paris VI),  France, in 2015. He is presently pursuing a Ph.D. in mathematics at Paris Diderot University (Paris VII),  France, under the supervision of Prof. Daniel Bennequin. 

 His work focuses on algebraic characterizations of information functions and related objects in probability theory and combinatorics, along with the application of homological and homotopical techniques in these domains. He is also interested in the relation between information theory and statistical mechanics, particularly the principle of free energy minimization and related algorithms (e.g. generalized belief propagation), as well as the use of this principle in machine learning and neuroscience. 
\end{IEEEbiographynophoto}

\end{document}